\documentclass[twocolumn,usenames,dvipsnames,table,xcdraw]{aastex631}
\pdfoutput=1
\usepackage{amsmath,amstext}
\usepackage[T1]{fontenc}
\usepackage{apjfonts} 
\usepackage[figure,figure*]{hypcap}
\usepackage{verbatim}
\usepackage{fancyhdr}
\DeclareUnicodeCharacter{2212}{-}

\newcommand{\code}[1]{\texttt{#1}}
\graphicspath{{./}{figures/}}
\newcommand{\targlong}{COOL~J1323+0343~}
\newcommand{\targlongns}{COOL~J1323+0343}
\newcommand{\targ}{CJ1323~}
\newcommand{\targnospace}{CJ1323}
\shorttitle{\targ}
\shortauthors{Sukay et al.}
\submitjournal{ApJ}
\begin{document}

\title{COOL-LAMPS II. Characterizing the Size and Star Formation History of a Bright Strongly Lensed Early-Type Galaxy at Redshift 1.02}

\author[0000-0002-1106-4881]{Ezra Sukay}
\affiliation{Department of Astronomy and Astrophysics, University of
Chicago, 5640 South Ellis Avenue, Chicago, IL 60637, USA}

\author[0000-0002-3475-7648]{Gourav Khullar}
\affiliation{Department of Astronomy and Astrophysics, University of
Chicago, 5640 South Ellis Avenue, Chicago, IL 60637, USA}
\affiliation{Kavli Institute for Cosmological Physics, University of
Chicago, 5640 South Ellis Avenue, Chicago, IL 60637, USA}

\author[0000-0003-1370-5010]{Michael D. Gladders}
\affiliation{Department of Astronomy and Astrophysics, University of
Chicago, 5640 South Ellis Avenue, Chicago, IL 60637, USA}
\affiliation{Kavli Institute for Cosmological Physics, University of
Chicago, 5640 South Ellis Avenue, Chicago, IL 60637, USA}

\author[0000-0002-7559-0864]{Keren Sharon}
\affiliation{Department of Astronomy, University of Michigan, 1085 South University Drive, Ann Arbor, MI 48109, USA}

\author[0000-0003-3266-2001]{Guillaume Mahler}
\affiliation{Department of Astronomy, University of Michigan, 1085 South University Drive, Ann Arbor, MI 48109, USA}
\affiliation{Department of Physics, Durham University, South Road, Durham DH1 3LE}
\author[0000-0003-4470-1696]{Kate Napier}
\affiliation{Department of Astronomy, University of Michigan, 1085 South University Drive, Ann Arbor, MI 48109, USA}

\author[0000-0001-7665-5079]{Lindsey E. Bleem}
\affiliation{Kavli Institute for Cosmological Physics, University of
Chicago, 5640 South Ellis Avenue, Chicago, IL 60637, USA}
\affiliation{Argonne National Laboratory, High-Energy Physics Division,
9700 S. Cass Avenue, Argonne, IL 60439, USA}
\author[0000-0003-2200-5606]{H{\AA}kon Dahle}
\affiliation{Institute of Theoretical Astrophysics, University of Oslo, P.O. Box 1029, Blindern, NO-0315 Oslo, Norway}

\author[0000-0001-5097-6755]{Michael K. Florian}
\affiliation{Steward Observatory, University of Arizona, 933 North Cherry Ave., Tucson, AZ 85721, USA}

\author[0000-0003-2294-4187]{Katya Gozman}
\affiliation{Department of Astronomy, University of Michigan, 1085 South University Drive, Ann Arbor, MI 48109, USA}

\author[0000-0003-1266-3445]{Jason J. Lin}
\affiliation{Department of Astronomy and Astrophysics, University of
Chicago, 5640 South Ellis Avenue, Chicago, IL 60637, USA}

\author[0000-0002-8397-8412]{Michael N. Martinez}
\affiliation{Department of Astronomy and Astrophysics, University of
Chicago, 5640 South Ellis Avenue, Chicago, IL 60637, USA}

\author[0000-0001-9225-972X]{Owen S. Matthews Acu\~{n}a}
\affiliation{Department of Astronomy and Astrophysics, University of
Chicago, 5640 South Ellis Avenue, Chicago, IL 60637, USA}

\author{Elisabeth Medina}
\affiliation{Department of Astronomy and Astrophysics, University of
Chicago, 5640 South Ellis Avenue, Chicago, IL 60637, USA}

\author[0000-0001-5931-5056]{Kaiya Merz}
\affiliation{Department of Astronomy and Astrophysics, University of
Chicago, 5640 South Ellis Avenue, Chicago, IL 60637, USA}

\author[0000-0002-9142-6378]{Jorge A. Sanchez}
\affiliation{Department of Astronomy and Astrophysics, University of
Chicago, 5640 South Ellis Avenue, Chicago, IL 60637, USA}

\author[0000-0002-2358-928X]{Emily E. Sisco}
\affiliation{Department of Astronomy and Astrophysics, University of
Chicago, 5640 South Ellis Avenue, Chicago, IL 60637, USA}

\author[0000-0001-8008-7270]{Daniel J. Kavin Stein}
\affiliation{Department of Astronomy and Astrophysics, University of
Chicago, 5640 South Ellis Avenue, Chicago, IL 60637, USA}

\author[0000-0001-6584-6144]{Kiyan Tavangar}
\affiliation{Department of Astronomy and Astrophysics, University of
Chicago, 5640 South Ellis Avenue, Chicago, IL 60637, USA}

\author[0000-0001-7160-3632]{Katherine E. Whitaker}
\affiliation{Department of Astronomy, University of Massachusetts, Amherst, MA 01003, USA}
\affiliation{Cosmic Dawn Center (DAWN), Denmark}

\email{Author for correspondence: sukay@uchicago.edu}

%\author{the COOL-LAMPS Collaboration}

%%%%%%%%%%%%%%%%%%%%%%%%%%%%%%%%%%%%%%%%%%%%%%%%%%%%%%%%%%%%%%%%%%%%%%%%%%%%%%%%%%%%%%%%%%
\begin{abstract}
We present COOL J1323+0343, an early-type galaxy at $z = 1.0153 \pm 0.0006$, strongly lensed by a cluster of galaxies at $z = 0.353 \pm 0.001$. This object was originally imaged by DECaLS and noted as a gravitational lens by COOL-LAMPS, a collaboration initiated to find strong-lensing systems in recent public optical imaging data, and confirmed with follow-up data. With ground-based $grzH$ imaging and optical spectroscopy from the Las Campanas Observatory and the Nordic Optical Telescope, we derive a stellar mass, metallicity, and star-formation history from stellar-population synthesis modeling. The lens modeling implies a total magnification of $\mu \sim $113. The median remnant stellar mass in the source plane is M$_* \sim 10.63$ $M_\odot$ and the median star-formation rate in the source plane is SFR $\sim 1.55 \times 10^{-3}$ M$_\odot$ yr$^{-1}$ (log sSFR = -13.4 yr$^{-1}$) in the youngest two age bins (0-100 Myr), closest to the epoch of observation. Our measurements place \targlong below the characteristic mass of the stellar mass function, making it an especially compelling target that could help clarify how intermediate mass quiescent galaxies evolve. We reconstruct \targlong in the source plane and fit its light profile. This object is below the expected size-evolution of early-type galaxy at this mass with an effective radius r$_e \sim$ 0.5 kpc. This extraordinarily magnified and bright lensed early-type galaxy offers an exciting opportunity to study the morphology and star formation history of an intermediate mass early-type galaxy in detail at $z \sim $1 .

\end{abstract}

%% Keywords should appear after the \end{abstract} command. 
%% See the online documentation for the full list of available subject
%% keywords and the rules for their use.

%%%%%%%%%%%%%%%%%%%%%%%%%%%%%%%%%%%%%%%%%%%%%%%%%%%%%%%%%%%%%%%%%%%%%%%%%%%%%%%%%%%%%%%%%%

\section{Introduction} \label{sec:intro}

Untangling the mechanisms that fuel the evolution of early-type galaxies (ETGs) is a key component to understanding how the diverse population of galaxies in the local Universe formed. The discovery that ETGs at $z>1$ are much more compact than those in the local Universe, with radii between 3-5 times smaller without much change in mass, is a challenge to our understanding of galaxy evolution (\citealp{Daddi_2005},  \citealp{Trujillo_2006}, \citealp{Trujillo_2007}, \citealp{VanDokkum_2008}, \citealp{Newman_2010}). There are two theories proposed to explain this growth. First, mergers with small, gas-poor satellite galaxies—known as minor mergers. Second, internal mechanisms like adiabatic expansion. Minor dry mergers would increase the radii of ETGs without requiring the addition of a proportional amount of mass (\citealp{Bezanson_2009}, \citealp{Naab_2009}) and some studies found that they are consistent with observations of ETGs at $z<1.6$ (\citealp{Belli_2014}). Minor mergers may explain the evolution seen at low redshifts, but they are insufficient to explain the rapid evolution and the scatter in radii at higher redshifts (\citealp{Fan_2010}, \citealp{Newman_2012}, \citealp{Nipoti_2012}). Furthermore, surveys have struggled to find the number of companion satellites required \citep{Newman_2012}. Adiabatic expansion triggered by AGN feedback, in combination with dry mergers, might resolve these inconsistencies (\citealp{Fan_2010}).

Another potential explanation is progenitor bias—the idea that the processes that quench ETGs at lower redshifts are different from those at $z > 2$ and, as a result, latecomers to the ETG population have larger radii. Number density studies found strong evidence that progenitor bias is not sufficient to explain the growth of massive ETGs (\citealp{Newman_2010}, \citealp{Belli_2014}, \citealp{Belli_2015}). For intermediate-mass ETGs (those with 10.5 < logM$_*$ < 11), some studies found that progenitor bias explains the majority of the observed evolution after $z = 1$ (\citealp{Carollo_2013}, \citealp{Fagioli_2016}). However, many other studies found evidence that individual growth is needed, at least in part, to explain the growth of intermediate-mass ETGs (\citealp{Cassata_2011}, \citealp{Newman_2012}, \citealp{Whitaker_2012}, \citealp{Belli_2014}, \citealp{vanderWel_2014}, \citealp{Belli_2015}).

The mechanisms that fuel the evolution of ETGs after they quench affects the morphology of evolving objects. Dry mergers with low mass objects result in central regions that have similar densities to very compact ETGs at $z \sim 2$, with an envelope of low-density material, high S\'ersic indices ($n \ge 5$), and negative metallicity gradients (\citealp{Hopkins_2009}, \citealp{Hilz_2013}). Adiabatic processes caused by quasar feedback would make the central regions of ETGs less dense in the local Universe than they are at $z \sim 2$ (\citealp{Fan_2010}). In-depth morphological studies will also provide clues to how ETGs quench in the first place. Simulations suggest galaxies that quenched "inside-out" through a central starburst have younger central stars than those on the edge \citep{Wellons_2015}. However, galaxies that quenched "outside-in" through cold gas accretion have stellar ages that are the same throughout or older central stars \citep{Feldman_2016}.

Spatially resolved imaging and spectroscopy targeting ETGs from $0.5<z<2$ should allow us to understand how they quench and what processes drive their structural evolution after star formation ceases. It is difficult to spatially resolve the most compact systems that may be little-modified descendants of compact high-redshift ETGs \citep[e.g.,][]{Stockton_2014}. Strong gravitational lensing enables study of more representative quiescent galaxies with better spatial resolution and signal-to-noise (e.g., \citealp{Oldham_2017}, \citealp{Akhshik_2020}, \citealp{Akhshik_2021}, \citealp{Man_2021}). Taking advantage of lensing magnification, \cite{Akhshik_2020,Akhshik_2021} were able to measure the age gradients and SFHs across seven spatial bins in a massive quiescent galaxy at $z = 1.88$. More examples of lensed ETGs, particularly with large magnifications that enable detailed studies, is key to further progress.

We discuss here the discovery and initial characterization of \targlongns: a compact intermediate mass ETG with old stellar populations which would be near impossible to study in-depth without strong gravitational lensing. \targlong extraordinary magnification of $\mu>100$ offers the opportunity to study a representative ETG at $z \sim 1$, and makes it an especially compelling target for more detailed follow-up imaging and spectroscopy. 

This paper is structured as follows. Section~\ref{sec:discovery} briefly describes the discovery of \targlongns. Section~\ref{sec:analysis} describes the follow-up imaging and spectroscopy of \targlongns. In Section~\ref{sec:stellarpops}, we report the results of stellar-population synthesis modeling. The lens modeling and source plane reconstruction are described in Section~\ref{sec:lensmodel}. Section~\ref{sec:con} puts \targlong in context with other ETGs and discusses what we expect to learn from detailed follow-up.

All reported magnitudes are calibrated to the AB system. The fiducial cosmology model used assumes a standard flat cold dark-matter model with a cosmological constant ($\Lambda$ CDM), corresponding to WMAP9 observations \citep{Hinshaw_2013}. For inferred parameters with uncertainties, we report 16th, 50th and 84th percentile values, unless otherwise specified.

\section{Discovery} 
\label{sec:discovery}

\targlong (hereafter \targnospace) (13h23m04.12s 03$^{\circ}$43'19.4") was discovered in a search for strong lenses in the Dark Energy Camera Legacy Survey (DECaLS) Data Release 8 (DR8, \citealp{Dey_2019}) by COOL-LAMPS - ChicagO Optically-selected strong Lenses - Located At the Margins of Public Surveys. Details about this search and results will be described in a future publication, so we only provide brief details here.

We found \targ as part of a comprehensive visual search of the northern galactic cap portion of the southern DECaLS dataset. Specifically, \targ was found in a search that examined lines of sight centered on luminous red galaxies, out to a photometric redshift of $z\sim0.7$. Both the foreground lensing cluster and the lensed source were immediately apparent in the DECaLS images; six co-authors examined this particular field yielding an aggregate score of 2.4 on a scale of 0 to 3, where a score of 3 signifies a definite lens.

The foreground lens was first noted as a cluster by \cite{Hao_2010} and was included in several other more recent cluster catalogs (e.g., \citealp{Rykoff_2014}, \citealp{Hilton_2021}). 
\targ was independently discovered as a strong lensing candidate by \cite{Huang_2021}, and noted as DESI-200.7678+03.7216. They first searched DR7 with a residual neural network (\citealp{Lanusse_2018}) using a training sample consisting of only observed images, but did not find this lens \citep{Huang_2020}. Using an improved "shielded" model with a larger training set on the complete DR8 dataset, the neural net presented in \cite{Huang_2021} gave \targ a probability of 0.34, well above their threshold of 0.1. Through visual inspection, it was assigned a final grade of 4 on a scale of 1 to 4, which indicates it is among the systems showing the clearest evidence of strong lensing.

\section{Follow-Up Observations and Analysis}
\label{sec:analysis}
\subsection{Imaging} \label{ssec:imaging}
\begin{figure}[t]

   {\includegraphics[width=0.47
   \textwidth]{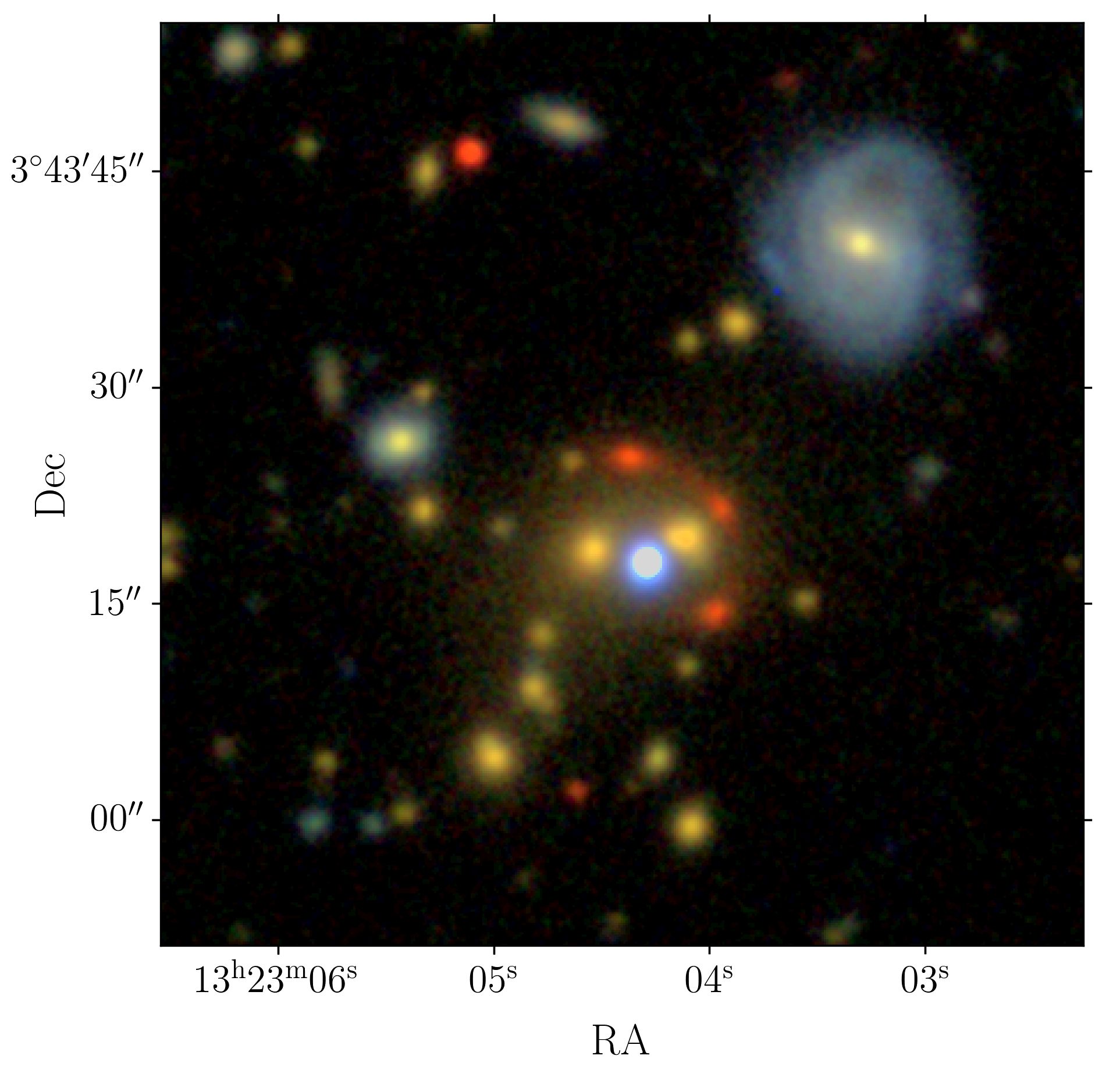}\label{fig:rgb+H}}

  \hfill
  \caption{$gr (z + H$) image of COOL J1323+0343. This image was constructed using $grz$-band imaging from DECaLS and $H$-band imaging from FOURSTAR on Magellan. The lensed galaxy is seen as a bright red multiply-imaged source, with three images visible in this figure. The foreground cluster has two similarly bright central galaxies. The bright blue source between the two brightest cluster galaxies is a foreground star.}
\end{figure}

Near-infrared imaging of \targ in the $H$-band was obtained using the FourStar Infrared Camera (FOURSTAR; \citealp{Persson_2008}) on the Magellan/Baade telescope, Chile, on February 22nd, 2020. The total integration time was 183s; the apparent brightness despite the briefness of this total time is a testament to the brightness of the lensed source images.
We reduced the data to final astrometrically- and photometrically-calibrated stacked images using a custom pipeline built via IRAF (\citealp{Tody1986}, \citealp{Tody1993}) and PHOTPIPE (\citealp{Rest_2005}, \citealp{Garg_2007}, \citealp{Miknaitis_2007}). We show a color image combining the $H$-band image with the DECaLS $grz$ data in Figure \ref{fig:rgb+H}.

The $H$-band image was calibrated to 2MASS stars \citep{Skrutskie_2006} within the field of view, with the calibration derived automatically by PHOTPIPE routines. Uncertainty on the zeropoint relative to 2MASS is $\sim 0.02$ magnitudes. We used the provided zeropoints for the DECaLs data; the uncertainties on these values were insignificant compared to other measurement uncertainties.

While the photometric analysis in this paper was completed primarily with the DECaLs optical imaging and FOURSTAR near-IR imaging, late in the preparation of this manuscript we obtained $grz$-band imaging. The images were taken with the Low Dispersion Survey Spectrograph (LDSS-3\footnote{\url{http://www.lco.cl/?epkb_post_type_1=ldss-3-user-manual}}) on the Magellan/Baade telescope, Chile, on January 16th, 2021, for a total integration time of 360s per filter, in subarcsecond seeing conditions. These higher resolution images were used to confirm the lensing configuration implied by the earlier data, as described in Section \ref{sec:lensmodel}.
%====================================================

%====================================================

\subsection{Spectroscopy}
Spectroscopic observations of \targ were obtained on April 19th, 2020, using the Alhambra Faint Object Spectrograph and Camera (ALFOSC) at the 2.56m Nordic Optical Telescope (NOT). Two 2400s exposures were obtained using Grism \#20 ($R=590$, $\lambda = 5650 - 10150$\AA), using a 1\farcs3 longslit. Halogen lamps were used for spectroscopic flat fielding, and wavelength solutions were calibrated using HeNe and ThAr arc lamps. Spectrophotometric calibration was performed using observations of the standard star SP1045$+$378. The longslit was placed to sample the most northern and most southern apparent images of the lensed source, and a dither along the slit was made between the two exposures sufficient to place both spectra of the source on slit regions that were clear of other objects. 

Reduction to one dimensional wavelength- and flux-calibrated spectra was accomplished using standard routines in IRAF. Sky-subtraction was ultimately accomplished using A-B subtraction of the two dithered spectra; while we explored sky-subtraction using adjacent sky apertures, we found they gave poor results. Light from the star and lens galaxy in between the two source images (see Figure \ref{fig:rgb+H}) contaminates regions that might otherwise be used as sky apertures and limits the accuracy with which the sky can be subtracted in this case.

\subsection{Redshifts}
 \targ is lensed by GMBCG J2007+03722 \citep{Hao_2010}, a galaxy cluster indicated by the abundance of red-sequence galaxies \citep{Gladders2000} easily visible in Figure \ref{fig:rgb+H} as elliptical galaxies with a similar orange hue. The two brightest of these galaxies near the cluster center have redshifts reported in the SDSS Legacy Survey at $z = 0.3535 \pm 0.0001$ \citep[]{York_2000,Strauss_2002} and by BOSS at $z = 0.35256 \pm 0.00006$ \citep[]{Eisenstein_2011,Dawson_2013} for the east and west galaxies, respectively. We took the cluster redshift as the average of these two values.

\begin{figure*}
    {\includegraphics[width=0.99
   \textwidth]{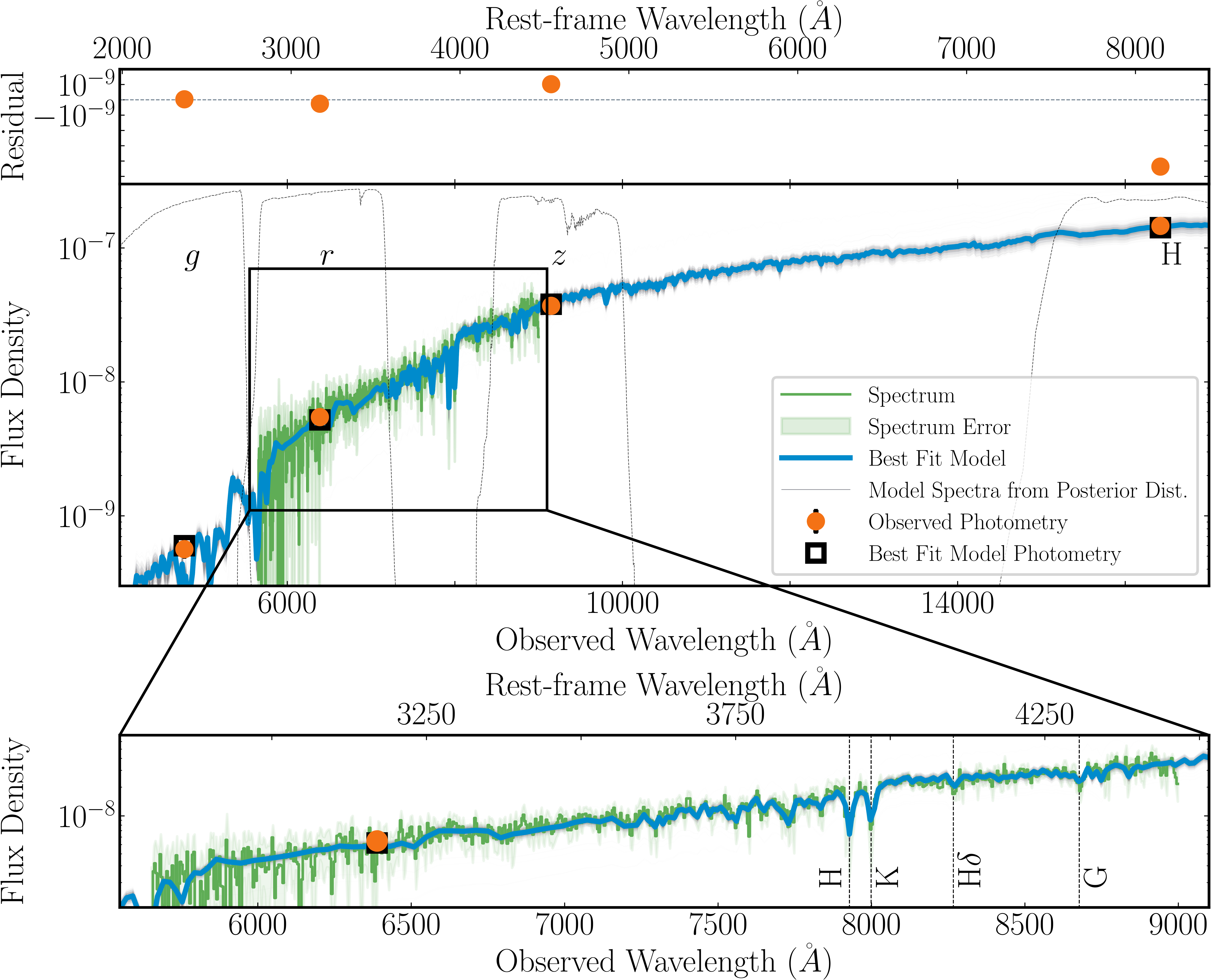}\label{fig:bestfit}}
  \hfill
  \caption{(Top) Residual values for best-fit photometry. (Middle) Best fit SED model using the \cite{Kriek_2013} dust attenuation curve (blue) and other fits from the posterior distribution (gray) calculated via Prospector using $grzH$ photometry (orange) and optical/NIR Nordic Optical Telescope/ALFOSC spectroscopy (green), with the spectra's $1\sigma$ error shown in light green. Best-fit photometry is shown as black squares. (Bottom) A zoom in on the black box in the middle plot to show the spectra in more detail, with H-delta emission line, Calcium H and K absorption lines, and the center of the G-band in the rest-frame labeled. All three plots are in units of maggies.}
\end{figure*}

The redshift of the lensed galaxy \targ is ${z = 1.0153 \pm 0.0006}$, based on Ca H \& K, H-$\delta$, and G-band features clearly visible in the NOT-ALFOSC spectrum shown in the bottom panel of Figure~\ref{fig:bestfit}.

\subsection{Model Photometry with GALFIT}
We used the parametric fitting code GALFIT (\citealp{Peng_2002,Peng_2010}) to fit light profiles to lensing cluster galaxies, other line-of-sight contaminants, and the targets of interest. For the $H$-band, we followed the process for making point-spread functions (PSFs) and utilizing GALFIT as described in \cite{Khullar_2021}. We used one or more 2D S\'ersic components to model the light of the components of the arc, galaxy cluster, and other objects. The foreground star between the cluster galaxies was fit with a PSF and an additional S\'ersic component to account for residual differences between this star and nearby isolated stars that were used to construct a reference PSF. Statistical magnitude uncertainties were measured as described in \cite{Khullar_2021}, with the final best fit model image injected into blank regions of the image, and then refit, with the distributions of results from these inject-and-recover tests providing the uncertainties. Additionally, models were built independently by three co-authors, and we found that the systematic uncertainties induced by the decision process inherent in this type of iterative model building was insignificant for the $H$-band measurements.

However, a similar initial analysis of the coarser DECaLS imaging indicated a significant issue with modeling systematics. Due to the overlapping mosaic of sampling from individual integrations that comprise the DECaLS images, there was no nearby isolated bright star to use as a reference, and hence in these data we were forced to use the star in the center of the field as a PSF reference. This was accomplished by first fitting the central region with a GALFIT model, absent any PSF convolution, with the star itself described by a two-component Moffat profile. The best fit model of the star was then extracted and used as a PSF reference for a complete GALFIT model of field. This limitation was possibly the source of the observed systematic differences between completed models. To investigate this further, we used GALFIT to fit nearby galaxies and stars with the targeted lensed images masked so that, after modeling, only the arc remained in the residual image. The photometry was then measured from this residual image using complex arc-like apertures at various scales, in order to measure the photometric curve of growth and a total lensed image magnitude \citep[e.g.,][]{Wuyts2010}. This second approach produced results consistent with other methods. In the following analysis, we used the resulting photometry from this method for the $grz$ filters.

We accounted for Galactic extinction by adjusting our photometry using the values reported in \citet{Schlafly_2011} as implemented by the NASA/IPAC Extragalactic Database's extinction calculator.\footnote{\href{https://ned.ipac.caltech.edu/extinction_calculator}{The NASA/IPAC Extragalactic Database (NED)} is funded by the National Aeronautics and Space Administration and operated by the California Institute of Technology.} Finally, zeropoint uncertainties and the statistical and systematic uncertainties estimated as above were combined in quadrature to compute total uncertainties for each measurement of each physical object. Table \ref{table:arcmags} shows total magnitudes for the sum of the three lensed images visible in Figure \ref{fig:rgb+H}.
%=======================================================
\newcommand{\datacaption}{\targ Photometry}
\newcommand{\datacomments}{Data in AB magnitudes. $grz$ band imaging from DECaLS. $H$-band imaging from Magellan/FOURSTAR infrared imager.}
\begin{deluxetable}{l ccccccc}
\tablecolumns{7}
\tablewidth{0pt}
%\tabletypesize{\scriptsize}
\tablecaption{\datacaption}
\tablehead{& $g$ & $r$ & $z$ & H}
\startdata
Arc  & 
23.12 $ \pm 0.15$ & 
20.66 $ \pm 0.10$ &
18.58 $ \pm 0.10$ &
17.09 $_{-0.09}^{+0.08}$ \\
\enddata
{\footnotesize \tablecomments{ \datacomments }}
\label{table:arcmags}
\vspace{-7mm}
\end{deluxetable}

%CG1 & 20.166 $_{-0.136}^{+0.121}$ &  17.845 $_{-0.231}^{+0.190}$ & 17.413 $_{-0.059}^{+0.056}$ & 15.975 $_{-0.207}^{+0.174}$ & \\
%CG2 &19.992 $_{-0.120}^{+0.108}$ & 17.532 $_{-0.119}^{+0.107}$ & 17.056 $_{-0.063}^{+0.059}$ & 16.510 $_{-0.045}^{+0.043}$ & \\
%Arc & 25.110 $_{-99.99}^{+99.99}$ & 23.197 $_{-0.358}^{+0.353}$ & 21.616 $_{-0.050}^{+0.049}$ & 20.472 $_{-0.048}^{+0.047}$ &  20.53 $_{-0.285}^{+0.273}$ &  20.556 $_{-0.087}^{+0.084}$\\
%Counterimage & 99.990 $_{-99.99}^{+99.99}$ & 99.990 $_{-99.99}^{+99.99}$ & 24.164 $_{-0.297}^{+0.233}$ & 22.705 $_{-0.127}^{+0.105}$ &  23.074 $_{-0.793}^{+0.453}$ &  23.774 $_{-1.545}^{+0.613}$\\
% \end{tabular}

% \begin{tabular}{p{1.8cm} p{2.3cm} p{2.3cm}  p{2.3cm} p{2.2cm}  p{2.2cm} p{2.3cm}  }
%  \multicolumn{7}{c}{Final Magnitudes} \\
%  \hline
%  \hline
%   & \textit{g} & \textit{r} & \textit{i} & \textit{z} & J & H\\
%  \hline

%CG 1                 & 24.773 $_{-0.122}^{+0.115}$ & 22.993 $_{-0.138}^{+0.122}$ & 22.008 $_{-0.078}^{+0.073}$ & 21.047 $_{-0.058}^{+0.055}$ & 20.029 $_{-0.069}^{+0.065}$ & 19.382 $_{-0.112}^{+0.103}$\\
% CG 2                 & 24.380 $_{-0.081}^{+0.075}$ & 22.032 $_{-0.103}^{+0.094}$ & 21.412 $_{-0.064}^{+0.068}$ & 19.996 $_{-0.073}^{+0.069}$ & 19.469 $_{-0.032}^{+0.031}$ & 19.332 $_{-0.065}^{+0.062}$\\
%{\footnotesize \tablecomments{ \derivcomments }}
%\tablenotes
%=======================================================
%====================================================
%===================================================

\section{Stellar Populations in \targ}
\label{sec:stellarpops}

We derived the properties of stellar populations in \targ by jointly fitting the spectra and photometry in the image plane with \code{Prospector}, an MCMC-based stellar population synthesis and parameter inference framework (\citealp{Conroy_2010}; \citealp{emceehammer}; \citealp{prospector}; \citealp{Leja_2017}). We assumed a non-parametric star formation history (SFH). We used seven age bins: $0-50$, $50-100$, $100-500$, $500-1000$, $1000-3000$, $3000-5000$, and $5000-5800$ Myr in lookback time, with 5800 Myr being the age of the Universe at $z = 1.015$. The age bins were represented by the parameters log(SFR ratios n), referring to the ratio of total star formation in adjacent time bins. These ratios were fit with a continuity prior (see \citealp{Leja_2019} for details). We ran our model with 1024 walkers, 1024 iterations, and a burn in = [8192, 4096, 2048, 1024, 512].

%====================================================
\begin{figure}[t]
  \includegraphics[width=0.49\textwidth]{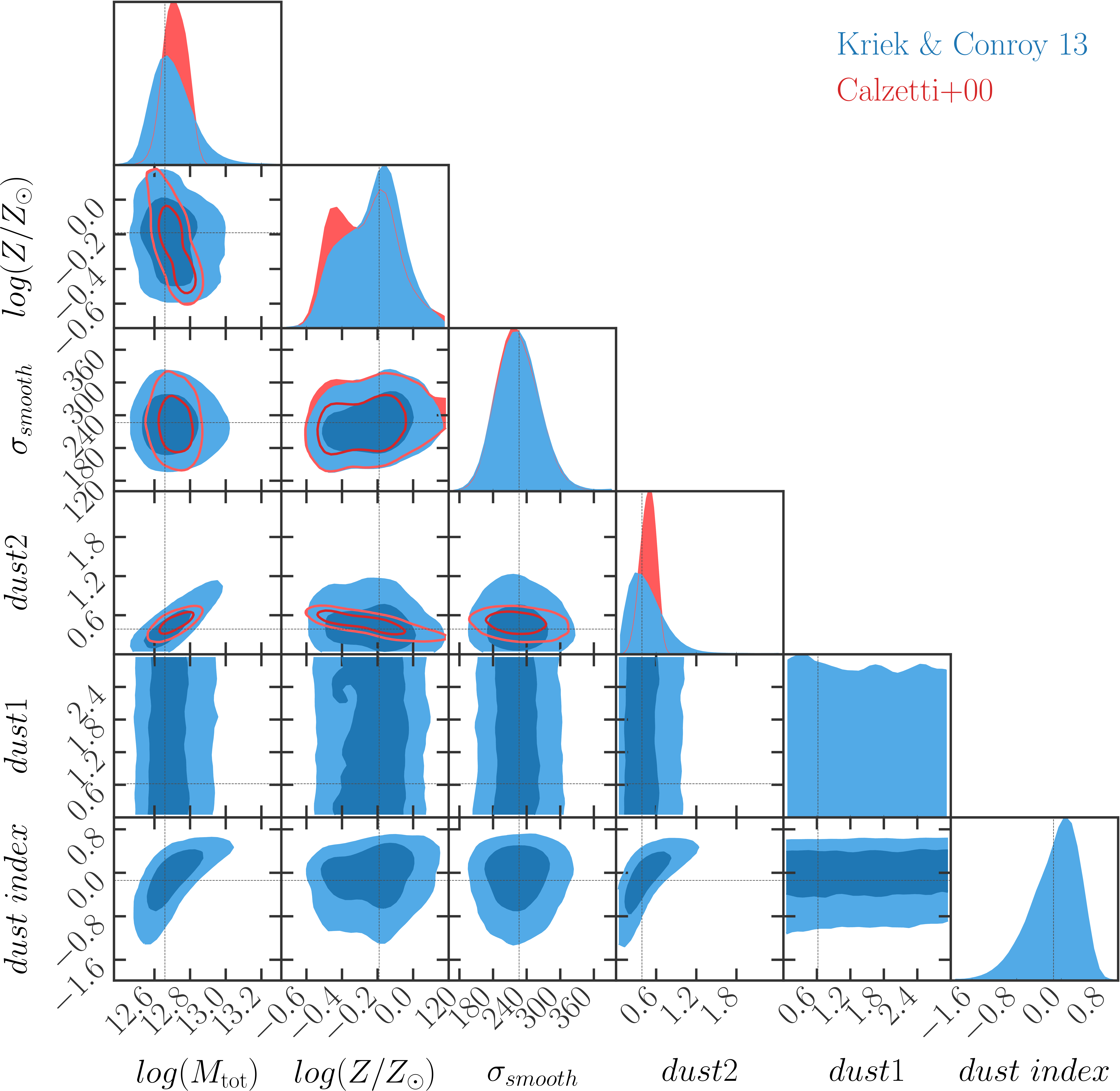}\label{fig:corner}
  \caption{Corner plot with posterior distributions and correlations for inferred parameters in the Prospector SED fitting analysis for the lensed source in the image plane, with contours corresponding to 1$\sigma$ (dark) and 2$\sigma$ (light). We show results from the models using the \cite{Kriek_2013} attenuation curve (blue) and the \cite{Calzetti_2000} attenuation curve (red). In the model using the C+00 curve, dust$\_$ index is fixed to 0 and dust2 is the only normalization factor. The best fit values for the model using the K\&C13 curve are shown with gray dashed lines. This plot clearly shows the results for mass and metallicity are similar regardless of model choice.}
\end{figure}
%====================================================
%====================================================
\begin{figure}
    \includegraphics[width=0.49\textwidth]{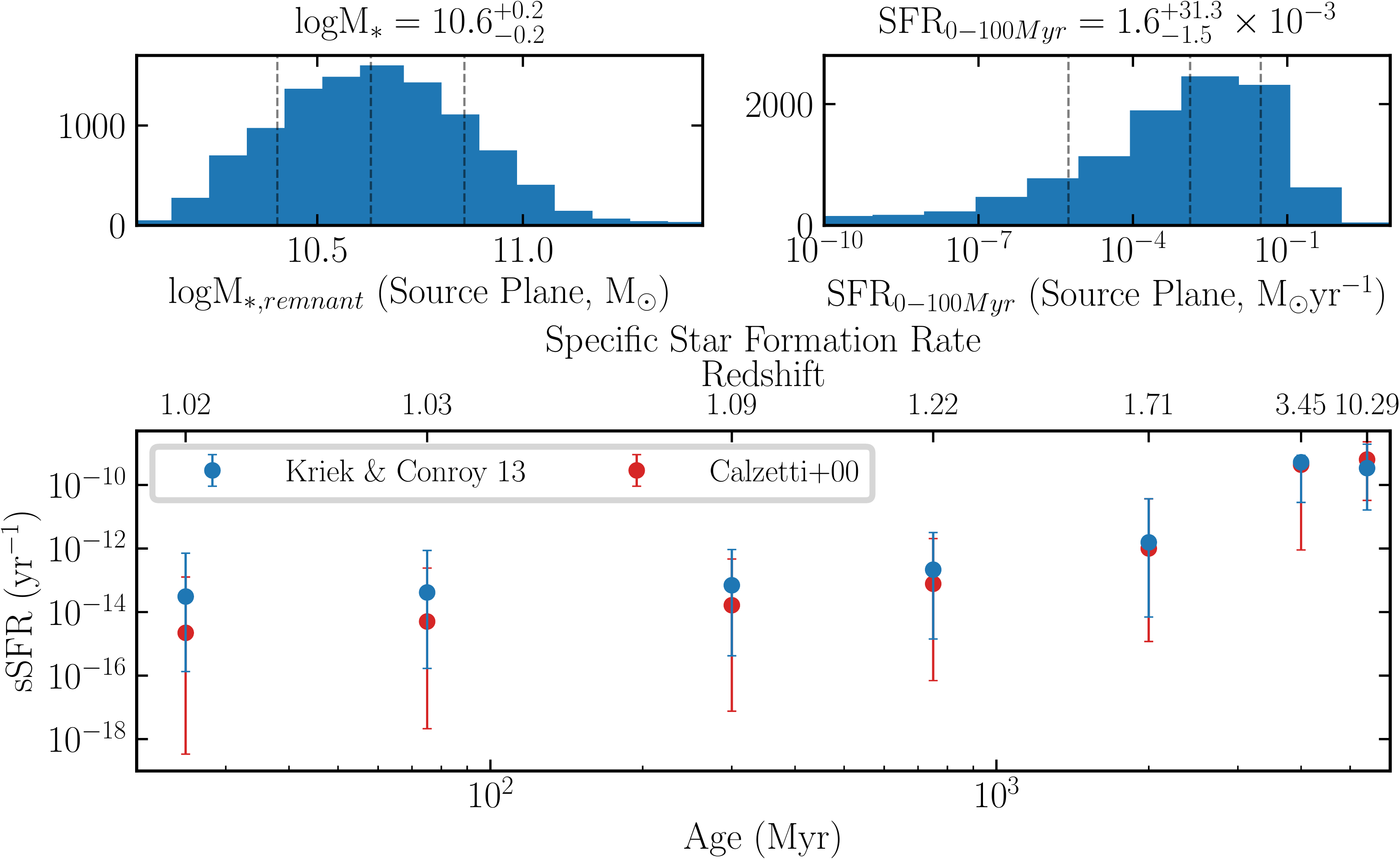}\label{fig:cornerssfrmass}
    \caption{(Top Left) The posterior distribution of the demagnified remnant stellar mass of \targ in the source plane. (Top Right) The posterior distribution of the star formation rate in the source plane. Both the remnant stellar mass and SFR are the results of the model utilizing the \cite{Kriek_2013} attenuation curve. The median and uncertainties are reported above each plot and shown as dashed lines. (Bottom) The specific star formation rate for each age bin. Both models show that \targ was quiescent at the epoch of observation and agree on the rate of star formation in early age bins. The model using the K\&C13 curve (blue) quenches faster and has a slight renewal in the most recent bin, while the model using the C+00 curve (red) quenches more gradually, with consistently lower sSFR in each age bin.} 
\end{figure}

\begin{figure*}

    {\includegraphics[width=0.5
   \textwidth]{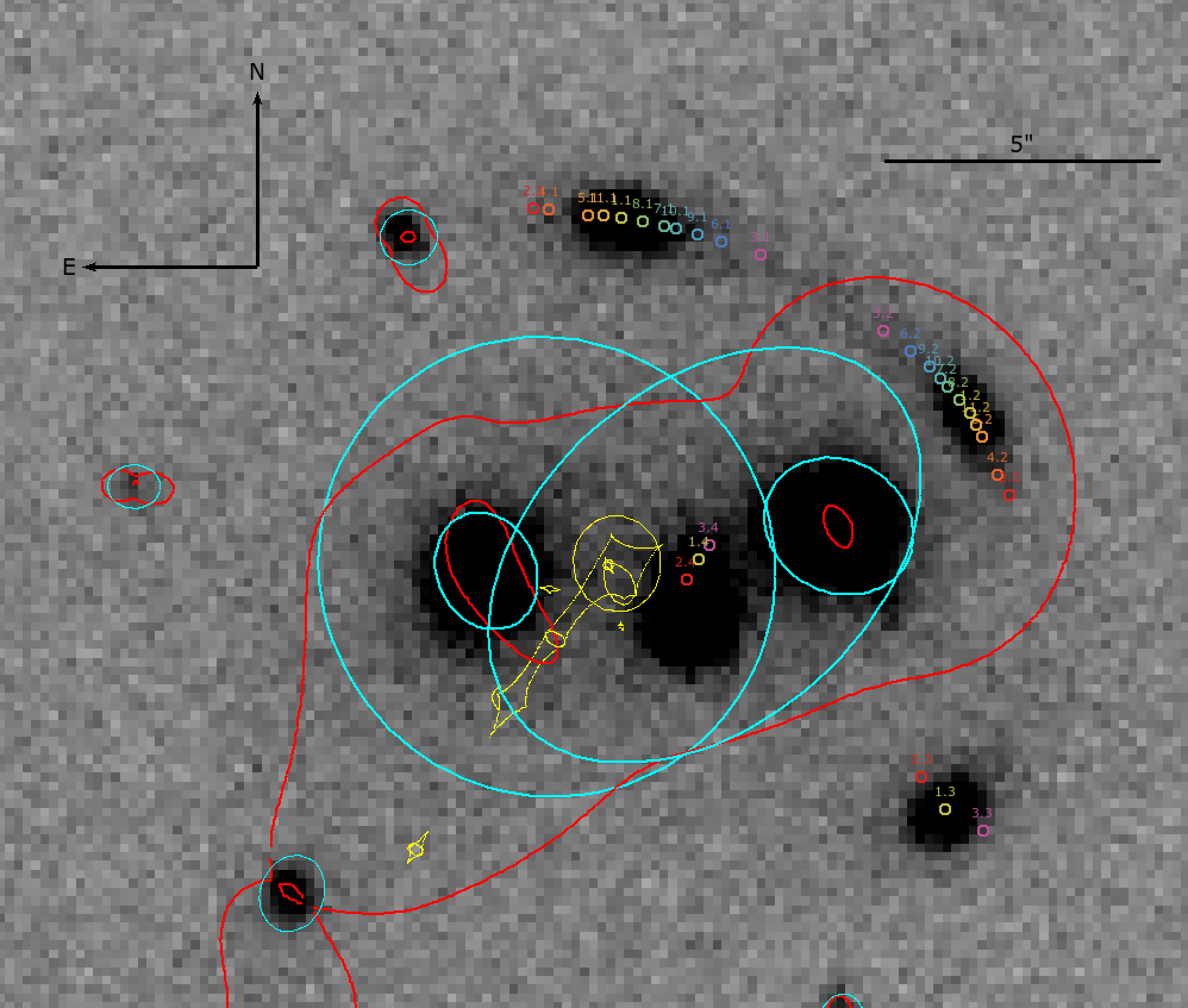}}
    {\includegraphics[width=0.5
   \textwidth]{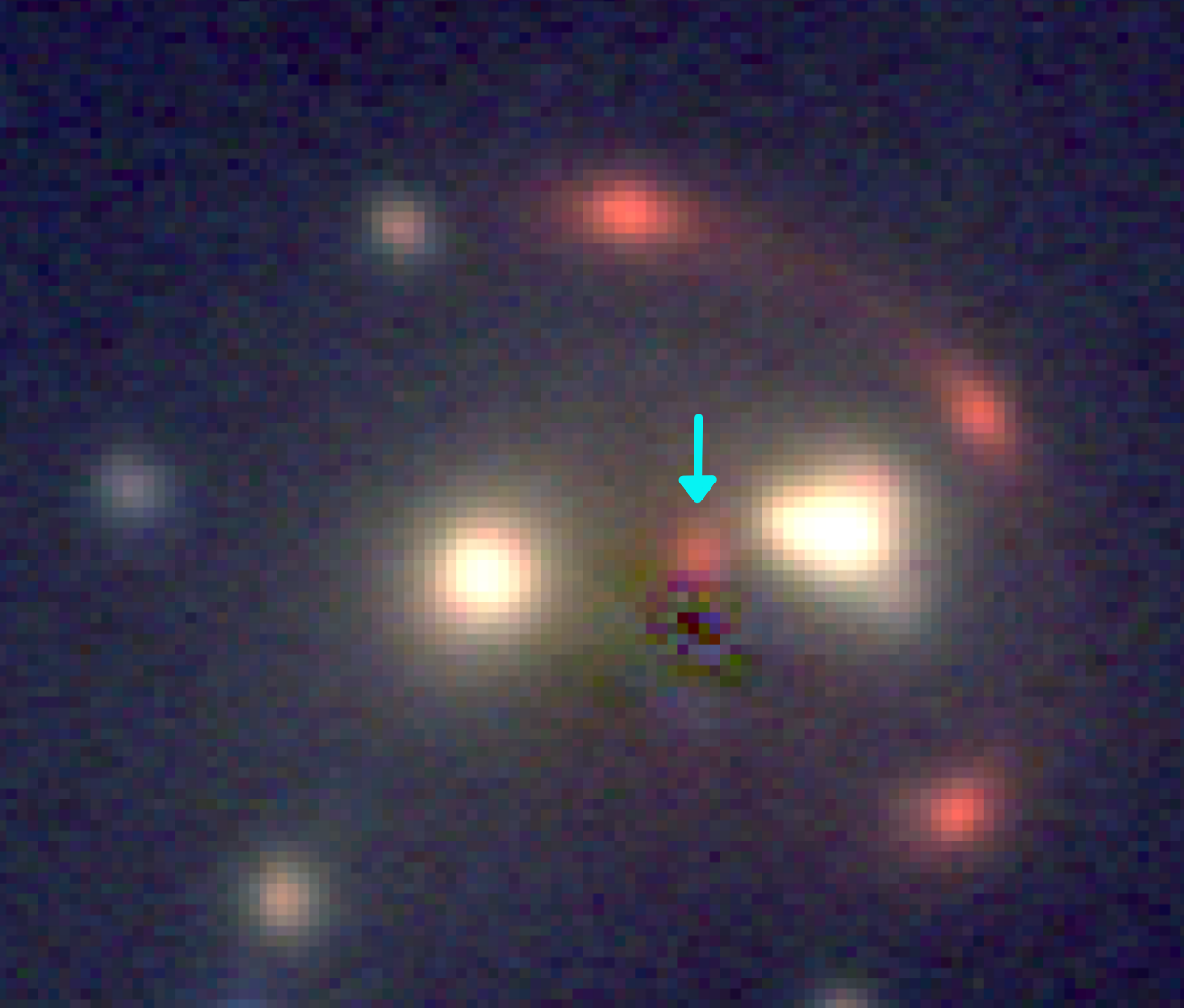}\label{fig:lensmodel}}
  \hfill
  \caption{(Left) FOURSTAR $H$-band image of \targ showing the tangential critical curve from the lens model (red) trisecting the arc and the caustic (yellow). The lens model was made using the positional constraints indicated with small circles. The positional constraints are derived from the $H$-band model, not convolved with the PSF, and each color represents locations of matching surface brightness on the lensed images.
  (Right) The LDSS3 $grz$-band image, with the central stellar contaminant subtracted, revealing a fourth image. The blue arrow points to the fourth image.}
\end{figure*}
%====================================================

We show corner plots constructed via \code{pyGTC} (\citealp{Bocquet2016}) for two different attenuation curves in Figure \ref{fig:corner}—the \cite{Kriek_2013} (K\&C13) attenuation curve is blue and the \cite{Calzetti_2000} (C+00) attenuation curve is red. We show the results of both models to give an idea of the systemic uncertainty introduced by how we model the dust attenuation. For both models, in addition to the star formation ratios, we simultaneously fit the following free parameters: total mass formed in the galaxy (M$_{tot}$, in units of M$_{\odot}$), stellar metallicity log$(Z/Z_{\odot})$ (where $Z_{\odot}=0.0142$), velocity smoothing in units of kms$^{-1}$, and spectrum normalization (the ratio of flux between the spectrum and photometry) to account for uncertainties in flux and spectral response calibration. For each chain in the MCMC analysis, we calculated the remnant stellar mass M$_{*,remnant}$. This accounted for mass loss from stars that have moved off the main-sequence, as well as stellar mass locked in stellar remnants, and is used as the mass throughout this paper. For the model using the C+00 attenuation curve, dust2 was a free parameter, which sets the overall normalization regardless of the age of stars, in units of opacity at 5500\AA. For the model using the K\&C13 attenuation curve, dust1 and dust2 were free parameters, which describe the attenuation of stellar light younger and older than $10^7$ years, respectively. Finally, for the K\&C13 model, dust$\_$index (dust slope) was a free parameter, which corresponds to the strength of the 2175 \AA \ UV bump. The stellar mass and metallicity distributions of both models are similar. However, the C+00 model's SFH suggests a slightly slower quenching than the K\&C13 model (see Figure \ref{fig:corner}). Furthermore, for the C+00 model, star formation declines up to the epoch of observation, while the K\&C13 model shows a slight rejuvenation in the most recent bin.

We show the best fit SED model using the K\&C13 dust attenuation curve in Figure \ref{fig:bestfit} and use the results of this model throughout the paper. The stellar mass and SFH of the arc are found to be robust when fixing stellar metallicity at the best fit value, log$(Z/Z_{\odot})$ = -0.19. The median value for the remnant stellar mass (M$_{*}$) in the image plane (i.e., the current value) is log(M/M$_{\odot}$) = 12.48$^{+0.12}_{-0.08}$ and the best fit SFR in the image plane is SFR = 1.98 $M_\odot$ yr$^{-1}$, averaged in the two youngest age bins (0-100 Myr). The median SFR = $0.11 ^{+2.28}_{-0.11} M_\odot$ yr$^{-1}$. (The mismatch between the median and best fit is due the SFR having a strong negative skew, as shown in the top right panel of Figure \ref{fig:cornerssfrmass}). We find that the best fit dust index is negative as seen in other quiescent and low mass galaxies (e.g., \citealp{Salim_2018}, \citealp{Whitaker_2021}), indicating that the attenuation curve is steeper than the \cite{Calzetti_2000} curve, where dust$\_$index = 0.

\section{Lens Modeling and Source Plane Reconstruction}
\label{sec:lensmodel}

\begin{figure}[!t]
  {\includegraphics[width=0.49\textwidth]{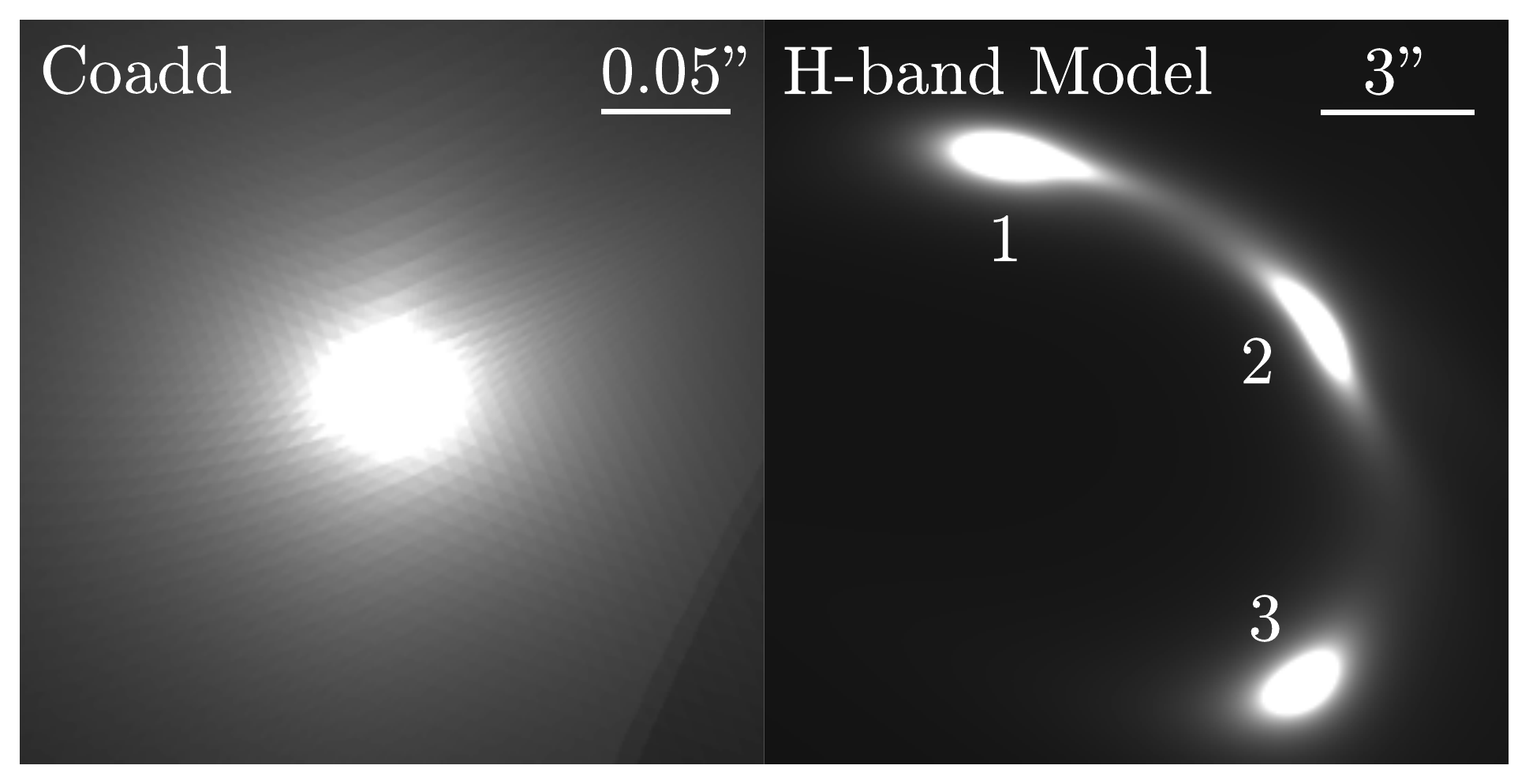}\label{fig:spr}}
  \hfill
  \caption{(Left) Coadded source plane reconstructions of COOL J1323+0343. (Right) The GALFIT model of the arc based on the FOURSTAR $H$-band data. The numbers next to the images correspond to the numbers used to refer to the images throughout the paper.}
  %h-band model: linear, vmin=-0.05,vmax=0.610644
  %Image 1-3: sqared, vmin=0, vmax=11.1194**2
\end{figure}

We modeled the mass distribution using LENSTOOL, a parametric lens modeling tool \citep{Jullo_2007}. All halos were treated as pseudo isothermal ellipsoids (dPIEs, \citealp{Eliasdottir_2007}) with parameters and uncertainties estimated using MCMC sampling. The overall modeling process closely followed the process described in \cite{Sharon_2020}. Astrometric constraints on the lens model were derived from the GALFIT model of the $H$-band image (the best seeing image available) without convolving the model with the PSF reference. This was essentially a parametric component deconvolution of the image. Astrometric reference points were placed on matching isophotes, which allowed us to use not just the centroid of each of the visible lensed images but also several locations along each image as constraints on the lens model. The appropriate configuration for these isophote-matched locations was guided by an initial simple model constructed using a single cluster-scale dPIE and smaller halos on each red-sequence cluster galaxy with positions, ellipticity, and position angles tied to the observed stellar light and the other parameters determined via scaling relations. Using the image labeling shown on the right panel of Figure \ref{fig:spr}, this initial model strongly suggests two key characteristics for this lensing configuration, namely that : 1) the primary magnification axis of image 3 is nearly orthogonal to images 1 and 2, and 2) that a fourth image should be present near the bright (and confounding) foreground star in the center of the field. While attempts to recover this fourth image from the initial DECaLS imaging were unsuccessful, additional optical imaging taken as described in Section \ref{ssec:imaging} does clearly show the expected fourth image, once the stellar image was fit and removed using GALFIT (see the right panel of Figure \ref{fig:lensmodel}).

The final lens model used four individually-described dPIE components—two fixed on the two visible bright central galaxies, and two cluster-scale halos allowed to vary about those positions. Other cluster-galaxy mass contributions were added via a scaling relation, as described above. The left panel of Figure \ref{fig:lensmodel} shows the caustic and critical curves for this final model, and the input constraints derived from the $H$-band image that informs the model.

With a final lens model in hand, we reconstructed \targ in the source plane with PyLenstool, a Python-based wrapper for LENSTOOL.\footnote{\url{http://pylenstool.readthedocs.io/}} The reconstructions were made with the un-convolved $H$-band GALFIT model as described above. We added together the source plane reconstructions, centered on the brightest point as determined by a single S\'ersic component GALFIT model. We show the combined image in the left panel of Figure \ref{fig:spr}. We then fit a S\'ersic and sky component to the coadded image using GALFIT and report the effective radius, S\'ersic index, and axis ratio in Table \ref{table:table2}. To understand the variation in the source plane morphology of \targ introduced by uncertainties in the model, we created combined images using the first $\sim$300 realizations of the lens model produced using the bayesCleanlens method in LENSTOOL. We report the median and uncertainty from the GALFIT models of these combined images in Table \ref{table:table2}. We find that all source plane reconstructions had $n < 3$ and $r_e < 1.7$, showing that our modeling is robust against degeneracies between $n$ and $r_e$.

%====================================================
%====================================================

We projected a contour of the image plane half light area into the source plane using the lens model. The resulting area in the source plane divided by the original image plane area is the magnification. The best fit magnifications are ${\mu_1 = 55}$, ${\mu_2 = 31}$, and ${\mu_3 = 28}$ for image 1, 2, and 3 respectively. The best fit total magnification is $\mu$ = 113 and the median is $\mu=74^{+49}_{-28}$. The magnification uncertainty was found using the same method on $\sim$ 300 realizations of the lens model. This gives a demagnified remnant median log$M_*$ = 10.63$^{+0.23}_{-0.23}$ $M_\odot$ and the demagnified best fit SFR = $1.75 \times 10^{-2}$ $M_\odot$ yr$^{-1}$ (from 0-100 Myr) (the median SFR = $1.55 ^{+31.31}_{-1.54} \times 10^{-3}$ $M_\odot$ yr$^{-1}$). 
We calculated the errors on the demagnified remnant mass and SFR by sampling the posterior distribution in concert with magnifications from realizations of the lens model. The results are displayed in the top left and right plots in Figure \ref{fig:cornerssfrmass}. 
%=======================================================
\newcommand{\arccaption}{Source Plane Morphology}
\newcommand{\arccomments}{The top row reports the GALFIT values for the coadd image shown in Figure \ref{fig:spr}. The second row reports the median with uncertainties from $\sim 300$ realizations of the lens model, as described in Section \ref{sec:lensmodel}. Radius reported in kpc.} 
\begin{deluxetable}{c c c c}

\tablecaption{\arccaption}

\tablehead{Image & $R_e$ & S\'ersic Index & Axis Ratio}
\startdata
Best Fit & 0.49 & 2.3 & 0.88 \\
Posterior Distribution & 0.58$^{+0.07}_{-0.09}$ & 2.2 $^{+0.1}_{-0.2}$ & 0.86 $^{+0.03}_{-0.03}$
\enddata
{\footnotesize \tablecomments{ \arccomments }}
\label{table:table2}
\vspace{-8mm}
\end{deluxetable}

%1 & 1.04 & 0.59 & 2.2 & 54.6 $\pm$ 16.3\\
%2 & 1.22 & 0.73 & 2.6 & 31.1 $\pm$ 9.2\\
%3  & 1.34 & 1.54 & 4.2 & 27.8 $\pm$ 9.3\\
%Combined & - & 0.49 & 2.3 & - 

%=======================================================

\section{Discussion and Future Work}
\label{sec:con}
\targ is a compact, intermediate mass quiescent galaxy at $z \sim $1—an object difficult to observe without gravitational lensing. Its stellar mass is $0.8^{+0.5}_{-0.3}$ times the characteristic mass of the stellar mass function reported in \citet{Muzzin_2013} for quiescent galaxies at $1 \le z < 1.5$. 
Compared with \citet{vanderWel_2014}'s expected size evolution for an ETG with M$_* = 4.27 \times 10^{10}$, \targ is small at 0.3 times the expected radius at its redshift, $\sigma=1$ below the relation. Furthermore, the stellar density of \targ is $\rho = 5^{+9}_{-3} \times 10^{10}$ $M_\odot$kpc$^{-3}$, comparable to compact ETGs at $z=2.3$ (e.g., \citealp{Kriek_2008}, \citealp{VanDokkum_2008}). This size is consistent with it being an unmodified relic ETG, that is, an early-type galaxy that quenched before $z = 2$ and has not grown since (e.g., \citealp{Stockton_2014}, \citealp{Hsu_2014}, \citealp{Trujillo_2014}, \citealp{FerreMateu_2017}).

The S\'ersic index of the combined source plane reconstructions is 2.3, indicating \targ is more disky than a standard $n = 4$ de Vaucouleurs profile. "Inside-out" growth leads disky ETGs at high redshifts to become elliptical in the local Universe as their outer envelopes build-up \citep{vanDokkum_2010}. Most samples of intermediate redshift ETGs found that a significant portion of their objects are disky (e.g., \citealp{Stockton_2010}, \citealp{Stockton_2014}, \citealp{Hsu_2014}).
However, all the source galaxies in \cite{Oldham_2017}'s survey of intermediate redshift, $0.4<z \le 0.7$, early-type/early-type lens systems (EELs) had at least one component with $n > 4$. That is, they have a bulge and are not disky. As \cite{Oldham_2017} noted, this could indicate that their targets are more evolved counterparts of objects like \targ and the ETGs in previous studies. We caution that source plane reconstructions based on ground based data are quite uncertain, and the preliminary results on morphology presented here should not be over-interpreted. Nevertheless, the suggestion from the apparent morphology and size is that \targ is at an early stage of its morphological evolution post-quenching. 

%====================================================
\begin{figure}[t]
  {\includegraphics[width=0.49\textwidth]{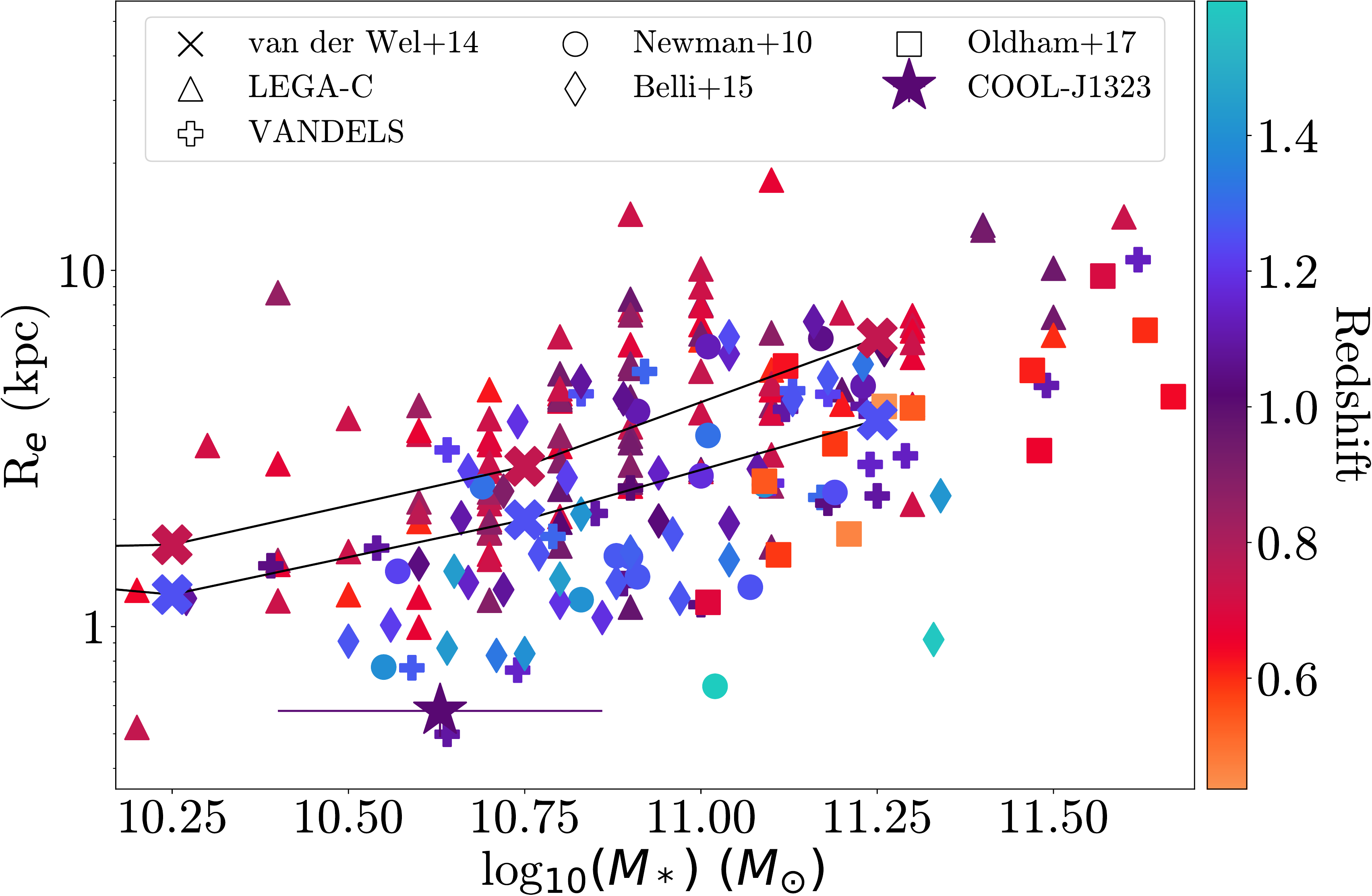}\label{fig:sizemass}}
  \hfill
  \caption{\citet{vanderWel_2014}'s size mass relation for ETGs at $z = 0.75$ and $z = 1.25$ shown with red and purple x's at the median, respectively, with black lines connecting each point. Hatched lines indicate the 16th and 86th percentiles. \targ is shown with a deep purple star. The color of each object corresponds to its redshift, as defined in the color bar to the right. The other symbols represent ETGs—triangles are quiescent galaxies at $z = 0.6-1$ reported in \cite{Bezanson_2018}, drawn from the LEGA-C survey (\citealp{vanderWel_2016}); plus-signs are 75 massive quiescent galaxies from $z = 1.0-1.3$ from the VANDELS survey \citep{Carnall_2019}; circles are 17 spheroidals at $z = 1.05-1.60$ from \cite{Newman_2010}, diamonds are 40 massive quiescent galaxies at $z = 1.0-1.6$ from \cite{Belli_2014} (the original sample is 56 galaxies—the 16 also in \cite{Newman_2010} have been removed), and squares are 13 strongly lensed massive compact ETGs at $z = 0.4-0.7$ from \cite{Oldham_2017}. It is clear that, on average, lower redshift galaxies have larger radii, despite there not being as significant a trend for mass. Furthermore, \targ is one of the most compact galaxies.} 
\end{figure} 
%====================================================

The SFH makes it clear that \targ has been quiescent for at least 1 Gyr before the epoch of observation. \cite{Man_2021}'s deep spectroscopic study found all three lensed quiescent galaxies had short bursts of star formation and then quenched quickly, with all three lying on the $\tau=0.1$ Gyr quenching curve from \cite{Belli_2019}. Artificial U-V and V-J colors (1.67 and 0.88, respectively) from the best fit SED model for \targ also lie on this curve. Deeper spectroscopy will reveal more about the SFH of \targnospace, including quenching time scale and age. While better resolution data is needed to fit a more granular SFH, \targnospace's SFH, S\'ersic index, radius, and stellar density suggest that it has not significantly changed since cosmic noon. If \targ is an unmodified relic, it will give us the opportunity to study the initial conditions of ETGs in exquisite detail. If not, future studies could uncover evidence for the post-quenching evolutionary mechanisms that fuel the growth in radii of ETGs.

The results from this preliminary data make it clear that \targ is a compelling target. Its total $H$-band magnitude of 17 makes it one of the two brightest lensed ETGs known, comparable only to MRG-M0138 (\citealp{Newman_2018}, \citealp{Jafariyazani_2020}). \targ can help probe the processes that cause ETG morphology to change so significantly. Resolved spectra—that are in principle possible because of \targnospace's extreme magnification—would clarify how it quenched by providing a more detailed and better resolved SFH, both spatially and temporally.
With sharper imaging, a more constrained lens model could be developed, yielding a more accurate and precise source-plane morphology of COOL J1323+0343, aided by near orthogonal magnifications provided by the different lensed images.

These observations are now being pursued for \targ and several other lensed ETGs discovered by the COOL-LAMPS collaboration.

\section*{Acknowledgements}

This work is supported by The College Undergraduate program at the University of Chicago, and the Department of Astronomy and Astrophysics at the University of Chicago. 

%Magellan/PISCO operations are supported by NSF AST1814719.
This paper includes data gathered with the 6.5 meter Magellan Telescopes located at Las Campanas Observatory, Chile.

%Standard acknowledgment texts for ALFOSC+NOT: 
Based in part on observations made with the Nordic Optical Telescope, operated by the Nordic Optical Telescope Scientific Association at the Observatorio del Roque de los Muchachos, La Palma, Spain, of the Instituto de Astrofisica de Canarias. The data presented here were obtained in part with ALFOSC, which is provided by the Instituto de Astrofisica de Andalucia (IAA) under a joint agreement with the University of Copenhagen and NOTSA.

%SDSS_II for left cluster galaxy 
Funding for the SDSS and SDSS-II has been provided by the Alfred P. Sloan Foundation, the Participating Institutions, the National Science Foundation, the U.S. Department of Energy, the National Aeronautics and Space Administration, the Japanese Monbukagakusho, the Max Planck Society, and the Higher Education Funding Council for England. The SDSS Web Site is \href{http://www.sdss.org/}{http://www.sdss.org/}.

The SDSS is managed by the Astrophysical Research Consortium for the Participating Institutions. The Participating Institutions are the American Museum of Natural History, Astrophysical Institute Potsdam, University of Basel, University of Cambridge, Case Western Reserve University, University of Chicago, Drexel University, Fermilab, the Institute for Advanced Study, the Japan Participation Group, Johns Hopkins University, the Joint Institute for Nuclear Astrophysics, the Kavli Institute for Particle Astrophysics and Cosmology, the Korean Scientist Group, the Chinese Academy of Sciences (LAMOST), Los Alamos National Laboratory, the Max-Planck-Institute for Astronomy (MPIA), the Max-Planck-Institute for Astrophysics (MPA), New Mexico State University, Ohio State University, University of Pittsburgh, University of Portsmouth, Princeton University, the United States Naval Observatory, and the University of Washington.

%the redshifts for the right cluster galaxy are from BOSS.
Funding for SDSS-III has been provided by the Alfred P. Sloan Foundation, the Participating Institutions, the National Science Foundation, and the U.S. Department of Energy Office of Science. The SDSS-III web site is \href{http://www.sdss3.org/}{http://www.sdss3.org/}.

SDSS-III is managed by the Astrophysical Research Consortium for the Participating Institutions of the SDSS-III Collaboration including the University of Arizona, the Brazilian Participation Group, Brookhaven National Laboratory, Carnegie Mellon University, University of Florida, the French Participation Group, the German Participation Group, Harvard University, the Instituto de Astrofisica de Canarias, the Michigan State/Notre Dame/JINA Participation Group, Johns Hopkins University, Lawrence Berkeley National Laboratory, Max Planck Institute for Astrophysics, Max Planck Institute for Extraterrestrial Physics, New Mexico State University, New York University, Ohio State University, Pennsylvania State University, University of Portsmouth, Princeton University, the Spanish Participation Group, University of Tokyo, University of Utah, Vanderbilt University, University of Virginia, University of Washington, and Yale University. 

%Legacy survey acknowledgement
The Legacy Surveys consist of three individual and complementary projects: the Dark Energy Camera Legacy Survey (DECaLS; NSF's OIR Lab Proposal ID \# 2014B-0404; PIs: David Schlegel and Arjun Dey), the Beijing-Arizona Sky Survey (BASS; NSF's OIR Lab Proposal ID \# 2015A-0801; PIs: Zhou Xu and Xiaohui Fan), and the Mayall z-band Legacy Survey (MzLS; NSF's OIR Lab Proposal ID \# 2016A-0453; PI: Arjun Dey). DECaLS, BASS and MzLS together include data obtained, respectively, at the Blanco telescope, Cerro Tololo Inter-American Observatory, The NSF's National Optical-Infrared Astronomy Research Laboratory (NSF's OIR Lab); the Bok telescope, Steward Observatory, University of Arizona; and the Mayall telescope, Kitt Peak National Observatory, NSF's OIR Lab. The Legacy Surveys project is honored to be permitted to conduct astronomical research on Iolkam Du'ag (Kitt Peak), a mountain with particular significance to the Tohono O'odham Nation.

The NSF's NOIR Lab is operated by the Association of Universities for Research in Astronomy (AURA) under a cooperative agreement with the National Science Foundation.

This project used data obtained with the Dark Energy Camera (DECam), which was constructed by the Dark Energy Survey (DES) collaboration. Funding for the DES Projects has been provided by the U.S. Department of Energy, the U.S. National Science Foundation, the Ministry of Science and Education of Spain, the Science and Technology Facilities Council of the United Kingdom, the Higher Education Funding Council for England, the National Center for Supercomputing Applications at the University of Illinois at Urbana-Champaign, the Kavli Institute of Cosmological Physics at the University of Chicago, Center for Cosmology and Astro-Particle Physics at the Ohio State University, the Mitchell Institute for Fundamental Physics and Astronomy at Texas A\&M University, Financiadora de Estudos e Projetos, Fundacao Carlos Chagas Filho de Amparo, Financiadora de Estudos e Projetos, Fundacao Carlos Chagas Filho de Amparo a Pesquisa do Estado do Rio de Janeiro, Conselho Nacional de Desenvolvimento Cientifico e Tecnologico and the Ministerio da Ciencia, Tecnologia e Inovacao, the Deutsche Forschungsgemeinschaft and the Collaborating Institutions in the Dark Energy Survey. The Collaborating Institutions are Argonne National Laboratory, the University of California at Santa Cruz, the University of Cambridge, Centro de Investigaciones Energeticas, Medioambientales y Tecnologicas-Madrid, the University of Chicago, University College London, the DES-Brazil Consortium, the University of Edinburgh, the Eidgenossische Technische Hochschule (ETH) Zurich, Fermi National Accelerator Laboratory, the University of Illinois at Urbana-Champaign, the Institut de Ciencies de l'Espai (IEEC/CSIC), the Institut de Fisica d'Altes Energies, Lawrence Berkeley National Laboratory, the Ludwig-Maximilians Universitat Munchen and the associated Excellence Cluster Universe, the University of Michigan, the National Optical Astronomy Observatory, the University of Nottingham, the Ohio State University, the University of Pennsylvania, the University of Portsmouth, SLAC National Accelerator Laboratory, Stanford University, the University of Sussex, and Texas A\&M University.

The Legacy Surveys imaging of the DESI footprint is supported by the Director, Office of Science, Office of High Energy Physics of the U.S. Department of Energy under Contract No. DE-AC02-05CH1123, by the National Energy Research Scientific Computing Center, a DOE Office of Science User Facility under the same contract; and by the U.S. National Science Foundation, Division of Astronomical Sciences under Contract No. AST-0950945 to NOAO.

%NED acknowledgement
This research has made use of the NASA/IPAC Extragalactic Database (NED), which is funded by the National Aeronautics and Space Administration and operated by the California Institute of Technology.

\facilities{CTIO/4m Blanco Telescope, Magellan Telescopes 6.5m (Baade/FOURSTAR, Clay/LDSS3), ALFOSC/2.56m Nordic Optical Telescope} 

\software{\code{Prospector, python-FSPS, SEDpy, pyGTC, Matplotlib \citep{matplotlib}, Numpy \citep{numpy}, Scipy \citep{scipy}, Astropy \citep{astropy}, LENSTOOL, PyLenstool, Jupyter, IPython Notebooks, GALFIT \citep{Peng_2002,Peng_2010}, SExtractor \citep{Bertin_1996}, SAO Image DS9 \citealp{Joye_2003}, IRAF \citep{Tody1986, Tody1993}}}

%% The reference list follows the main body and any appendices.
%% Use LaTeX's thebibliography environment to mark up your reference list.
%% Note \begin{thebibliography} is followed by an empty set of
%% curly braces.  If you forget this, LaTeX will generate the error
%% "Perhaps a missing \item?".
%%
%% thebibliography produces citations in the text using \bibitem-\cite
%% cross-referencing. Each reference is preceded by a
%% \bibitem command that defines in curly braces the KEY that corresponds
%% to the KEY in the \cite commands (see the first section above).
%% Make sure that you provide a unique KEY for every \bibitem or else the
%% paper will not LaTeX. The square brackets should contain
%% the citation text that LaTeX will insert in
%% place of the \cite commands.

%% We have used macros to produce journal name abbreviations.
%% \aastex provides a number of these for the more frequently-cited journals.
%% See the Author Guide for a list of them.

%% Note that the style of the \bibitem labels (in []) is slightly
%% different from previous examples.  The natbib system solves a host
%% of citation expression problems, but it is necessary to clearly
%% delimit the year from the author name used in the citation.
%% See the natbib documentation for more details and options.

\bibliographystyle{yahapj}
\bibliography{references1}

\begin{thebibliography}{}
\providecommand\natexlab[1]{#1}
\providecommand\JournalTitle[1]{#1}

\bibitem[{{Akhshik} {et~al.}(2020){Akhshik}, {Whitaker}, {Brammer}, {Mahler},
  {Sharon}, {Leja}, {Bayliss}, {Bezanson}, {Gladders}, {Man}, {Nelson},
  {Rigby}, {Rizzo}, {Toft}, {Wellons}, \& {Williams}}]{Akhshik_2020}
{Akhshik}, M., {Whitaker}, K.~E., {Brammer}, G., {et~al.} 2020,
  \href{http://dx.doi.org/10.3847/1538-4357/abac62}{\JournalTitle{\apj}, 900,
  184}

\bibitem[{{Akhshik} {et~al.}(2021){Akhshik}, {Whitaker}, {Leja}, {Mahler},
  {Sharon}, {Brammer}, {Toft}, {Bezanson}, {Man}, {Nelson}, {Pacifici},
  {Wellons}, \& {Williams}}]{Akhshik_2021}
{Akhshik}, M., {Whitaker}, K.~E., {Leja}, J., {et~al.} 2021,
  \href{http://dx.doi.org/10.3847/2041-8213/abd416}{\JournalTitle{\apjl}, 907,
  L8}

\bibitem[{{Astropy Collaboration} {et~al.}(2013){Astropy Collaboration},
  {Robitaille}, {Tollerud}, {Greenfield}, {Droettboom}, {Bray}, {Aldcroft},
  {Davis}, {Ginsburg}, {Price-Whelan}, {Kerzendorf}, {Conley}, {Crighton},
  {Barbary}, {Muna}, {Ferguson}, {Grollier}, {Parikh}, {Nair}, {Unther},
  {Deil}, {Woillez}, {Conseil}, {Kramer}, {Turner}, {Singer}, {Fox}, {Weaver},
  {Zabalza}, {Edwards}, {Azalee Bostroem}, {Burke}, {Casey}, {Crawford},
  {Dencheva}, {Ely}, {Jenness}, {Labrie}, {Lim}, {Pierfederici}, {Pontzen},
  {Ptak}, {Refsdal}, {Servillat}, \& {Streicher}}]{astropy}
{Astropy Collaboration}, {Robitaille}, T.~P., {Tollerud}, E.~J., {et~al.} 2013,
  \href{http://dx.doi.org/10.1051/0004-6361/201322068}{\JournalTitle{\aap},
  558, A33}

\bibitem[{{Belli} {et~al.}(2014){Belli}, {Newman}, \& {Ellis}}]{Belli_2014}
{Belli}, S., {Newman}, A.~B., \& {Ellis}, R.~S. 2014,
  \href{http://dx.doi.org/10.1088/0004-637X/783/2/117}{\JournalTitle{\apj},
  783, 117}

\bibitem[{{Belli} {et~al.}(2015){Belli}, {Newman}, \& {Ellis}}]{Belli_2015}
---. 2015,
  \href{http://dx.doi.org/10.1088/0004-637X/799/2/206}{\JournalTitle{\apj},
  799, 206}

\bibitem[{{Belli} {et~al.}(2019){Belli}, {Newman}, \& {Ellis}}]{Belli_2019}
---. 2019,
  \href{http://dx.doi.org/10.3847/1538-4357/ab07af}{\JournalTitle{\apj}, 874,
  17}

\bibitem[{{Bertin} \& {Arnouts}(1996)}]{Bertin_1996}
{Bertin}, E., \& {Arnouts}, S. 1996,
  \href{http://dx.doi.org/10.1051/aas:1996164}{\JournalTitle{\aaps}, 117, 393}

\bibitem[{{Bezanson} {et~al.}(2009){Bezanson}, {van Dokkum}, {Tal},
  {Marchesini}, {Kriek}, {Franx}, \& {Coppi}}]{Bezanson_2009}
{Bezanson}, R., {van Dokkum}, P.~G., {Tal}, T., {et~al.} 2009,
  \href{http://dx.doi.org/10.1088/0004-637X/697/2/1290}{\JournalTitle{\apj},
  697, 1290}

\bibitem[{{Bezanson} {et~al.}(2018){Bezanson}, {van der Wel}, {Pacifici},
  {Noeske}, {Bari{\v{s}}i{\'c}}, {Bell}, {Brammer}, {Calhau}, {Chauke}, {van
  Dokkum}, {Franx}, {Gallazzi}, {van Houdt}, {Labb{\'e}}, {Maseda},
  {Mu{\~n}os-Mateos}, {Muzzin}, {van de Sande}, {Sobral}, {Straatman}, \&
  {Wu}}]{Bezanson_2018}
{Bezanson}, R., {van der Wel}, A., {Pacifici}, C., {et~al.} 2018,
  \href{http://dx.doi.org/10.3847/1538-4357/aabc55}{\JournalTitle{\apj}, 858,
  60}

\bibitem[{Bocquet \& Carter(2016)}]{Bocquet2016}
Bocquet, S., \& Carter, F.~W. 2016,
  \href{http://dx.doi.org/10.21105/joss.00046}{\JournalTitle{The Journal of
  Open Source Software}, 1}

\bibitem[{{Calzetti} {et~al.}(2000){Calzetti}, {Armus}, {Bohlin}, {Kinney},
  {Koornneef}, \& {Storchi-Bergmann}}]{Calzetti_2000}
{Calzetti}, D., {Armus}, L., {Bohlin}, R.~C., {et~al.} 2000,
  \href{http://dx.doi.org/10.1086/308692}{\JournalTitle{\apj}, 533, 682}

\bibitem[{{Carnall} {et~al.}(2019){Carnall}, {McLure}, {Dunlop}, {Cullen},
  {McLeod}, {Wild}, {Johnson}, {Appleby}, {Dav{\'e}}, {Amorin}, {Bolzonella},
  {Castellano}, {Cimatti}, {Cucciati}, {Gargiulo}, {Garilli}, {Marchi},
  {Pentericci}, {Pozzetti}, {Schreiber}, {Talia}, \& {Zamorani}}]{Carnall_2019}
{Carnall}, A.~C., {McLure}, R.~J., {Dunlop}, J.~S., {et~al.} 2019,
  \href{http://dx.doi.org/10.1093/mnras/stz2544}{\JournalTitle{\mnras}, 490,
  417}

\bibitem[{{Carollo} {et~al.}(2013){Carollo}, {Bschorr}, {Renzini}, {Lilly},
  {Capak}, {Cibinel}, {Ilbert}, {Onodera}, {Scoville}, {Cameron}, {Mobasher},
  {Sanders}, \& {Taniguchi}}]{Carollo_2013}
{Carollo}, C.~M., {Bschorr}, T.~J., {Renzini}, A., {et~al.} 2013,
  \href{http://dx.doi.org/10.1088/0004-637X/773/2/112}{\JournalTitle{\apj},
  773, 112}

\bibitem[{{Cassata} {et~al.}(2011){Cassata}, {Giavalisco}, {Guo}, {Renzini},
  {Ferguson}, {Koekemoer}, {Salimbeni}, {Scarlata}, {Grogin}, {Conselice},
  {Dahlen}, {Lotz}, {Dickinson}, \& {Lin}}]{Cassata_2011}
{Cassata}, P., {Giavalisco}, M., {Guo}, Y., {et~al.} 2011,
  \href{http://dx.doi.org/10.1088/0004-637X/743/1/96}{\JournalTitle{\apj}, 743,
  96}

\bibitem[{{Conroy} \& {Gunn}(2010)}]{Conroy_2010}
{Conroy}, C., \& {Gunn}, J.~E. 2010,
  \href{http://dx.doi.org/10.1088/0004-637X/712/2/833}{\JournalTitle{\apj},
  712, 833}

\bibitem[{{Daddi} {et~al.}(2005){Daddi}, {Renzini}, {Pirzkal}, {Cimatti},
  {Malhotra}, {Stiavelli}, {Xu}, {Pasquali}, {Rhoads}, {Brusa}, {di Serego
  Alighieri}, {Ferguson}, {Koekemoer}, {Moustakas}, {Panagia}, \&
  {Windhorst}}]{Daddi_2005}
{Daddi}, E., {Renzini}, A., {Pirzkal}, N., {et~al.} 2005,
  \href{http://dx.doi.org/10.1086/430104}{\JournalTitle{\apj}, 626, 680}

\bibitem[{{Dawson} {et~al.}(2013){Dawson}, {Schlegel}, {Ahn}, {Anderson},
  {Aubourg}, {Bailey}, {Barkhouser}, {Bautista}, {Beifiori}, {Berlind},
  {Bhardwaj}, {Bizyaev}, {Blake}, {Blanton}, {Blomqvist}, {Bolton}, {Borde},
  {Bovy}, {Brandt}, {Brewington}, {Brinkmann}, {Brown}, {Brownstein}, {Bundy},
  {Busca}, {Carithers}, {Carnero}, {Carr}, {Chen}, {Comparat}, {Connolly},
  {Cope}, {Croft}, {Cuesta}, {da Costa}, {Davenport}, {Delubac}, {de Putter},
  {Dhital}, {Ealet}, {Ebelke}, {Eisenstein}, {Escoffier}, {Fan}, {Filiz Ak},
  {Finley}, {Font-Ribera}, {G{\'e}nova-Santos}, {Gunn}, {Guo}, {Haggard},
  {Hall}, {Hamilton}, {Harris}, {Harris}, {Ho}, {Hogg}, {Holder}, {Honscheid},
  {Huehnerhoff}, {Jordan}, {Jordan}, {Kauffmann}, {Kazin}, {Kirkby}, {Klaene},
  {Kneib}, {Le Goff}, {Lee}, {Long}, {Loomis}, {Lundgren}, {Lupton}, {Maia},
  {Makler}, {Malanushenko}, {Malanushenko}, {Mandelbaum}, {Manera}, {Maraston},
  {Margala}, {Masters}, {McBride}, {McDonald}, {McGreer}, {McMahon}, {Mena},
  {Miralda-Escud{\'e}}, {Montero-Dorta}, {Montesano}, {Muna}, {Myers},
  {Naugle}, {Nichol}, {Noterdaeme}, {Nuza}, {Olmstead}, {Oravetz}, {Oravetz},
  {Owen}, {Padmanabhan}, {Palanque-Delabrouille}, {Pan}, {Parejko},
  {P{\^a}ris}, {Percival}, {P{\'e}rez-Fournon}, {P{\'e}rez-R{\`a}fols},
  {Petitjean}, {Pfaffenberger}, {Pforr}, {Pieri}, {Prada}, {Price-Whelan},
  {Raddick}, {Rebolo}, {Rich}, {Richards}, {Rockosi}, {Roe}, {Ross}, {Ross},
  {Rossi}, {Rubi{\~n}o-Martin}, {Samushia}, {S{\'a}nchez}, {Sayres}, {Schmidt},
  {Schneider}, {Sc{\'o}ccola}, {Seo}, {Shelden}, {Sheldon}, {Shen}, {Shu},
  {Slosar}, {Smee}, {Snedden}, {Stauffer}, {Steele}, {Strauss}, {Streblyanska},
  {Suzuki}, {Swanson}, {Tal}, {Tanaka}, {Thomas}, {Tinker}, {Tojeiro},
  {Tremonti}, {Vargas Maga{\~n}a}, {Verde}, {Viel}, {Wake}, {Watson}, {Weaver},
  {Weinberg}, {Weiner}, {West}, {White}, {Wood-Vasey}, {Yeche}, {Zehavi},
  {Zhao}, \& {Zheng}}]{Dawson_2013}
{Dawson}, K.~S., {Schlegel}, D.~J., {Ahn}, C.~P., {et~al.} 2013,
  \href{http://dx.doi.org/10.1088/0004-6256/145/1/10}{\JournalTitle{\aj}, 145,
  10}

\bibitem[{{Dey} {et~al.}(2019){Dey}, {Schlegel}, {Lang}, {Blum}, {Burleigh},
  {Fan}, {Findlay}, {Finkbeiner}, {Herrera}, {Juneau}, {Landriau}, {Levi},
  {McGreer}, {Meisner}, {Myers}, {Moustakas}, {Nugent}, {Patej}, {Schlafly},
  {Walker}, {Valdes}, {Weaver}, {Y{\`e}che}, {Zou}, {Zhou}, {Abareshi},
  {Abbott}, {Abolfathi}, {Aguilera}, {Alam}, {Allen}, {Alvarez}, {Annis},
  {Ansarinejad}, {Aubert}, {Beechert}, {Bell}, {BenZvi}, {Beutler}, {Bielby},
  {Bolton}, {Brice{\~n}o}, {Buckley-Geer}, {Butler}, {Calamida}, {Carlberg},
  {Carter}, {Casas}, {Castander}, {Choi}, {Comparat}, {Cukanovaite}, {Delubac},
  {DeVries}, {Dey}, {Dhungana}, {Dickinson}, {Ding}, {Donaldson}, {Duan},
  {Duckworth}, {Eftekharzadeh}, {Eisenstein}, {Etourneau}, {Fagrelius},
  {Farihi}, {Fitzpatrick}, {Font-Ribera}, {Fulmer}, {G{\"a}nsicke},
  {Gaztanaga}, {George}, {Gerdes}, {Gontcho}, {Gorgoni}, {Green}, {Guy},
  {Harmer}, {Hernand ez}, {Honscheid}, {Huang}, {James}, {Jannuzi}, {Jiang},
  {Joyce}, {Karcher}, {Karkar}, {Kehoe}, {Kneib}, {Kueter-Young}, {Lan},
  {Lauer}, {Le Guillou}, {Le Van Suu}, {Lee}, {Lesser}, {Perreault Levasseur},
  {Li}, {Mann}, {Marshall}, {Mart{\'\i}nez-V{\'a}zquez}, {Martini}, {du Mas des
  Bourboux}, {McManus}, {Meier}, {M{\'e}nard}, {Metcalfe},
  {Mu{\~n}oz-Guti{\'e}rrez}, {Najita}, {Napier}, {Narayan}, {Newman}, {Nie},
  {Nord}, {Norman}, {Olsen}, {Paat}, {Palanque-Delabrouille}, {Peng},
  {Poppett}, {Poremba}, {Prakash}, {Rabinowitz}, {Raichoor}, {Rezaie},
  {Robertson}, {Roe}, {Ross}, {Ross}, {Rudnick}, {Safonova}, {Saha},
  {S{\'a}nchez}, {Savary}, {Schweiker}, {Scott}, {Seo}, {Shan}, {Silva},
  {Slepian}, {Soto}, {Sprayberry}, {Staten}, {Stillman}, {Stupak}, {Summers},
  {Sien Tie}, {Tirado}, {Vargas-Maga{\~n}a}, {Vivas}, {Wechsler}, {Williams},
  {Yang}, {Yang}, {Yapici}, {Zaritsky}, {Zenteno}, {Zhang}, {Zhang}, {Zhou}, \&
  {Zhou}}]{Dey_2019}
{Dey}, A., {Schlegel}, D.~J., {Lang}, D., {et~al.} 2019,
  \href{http://dx.doi.org/10.3847/1538-3881/ab089d}{\JournalTitle{\aj}, 157,
  168}

\bibitem[{{Eisenstein} {et~al.}(2011){Eisenstein}, {Weinberg}, {Agol},
  {Aihara}, {Allende Prieto}, {Anderson}, {Arns}, {Aubourg}, {Bailey},
  {Balbinot}, {Barkhouser}, {Beers}, {Berlind}, {Bickerton}, {Bizyaev},
  {Blanton}, {Bochanski}, {Bolton}, {Bosman}, {Bovy}, {Brandt}, {Breslauer},
  {Brewington}, {Brinkmann}, {Brown}, {Brownstein}, {Burger}, {Busca},
  {Campbell}, {Cargile}, {Carithers}, {Carlberg}, {Carr}, {Chang}, {Chen},
  {Chiappini}, {Comparat}, {Connolly}, {Cortes}, {Croft}, {Cunha}, {da Costa},
  {Davenport}, {Dawson}, {De Lee}, {Porto de Mello}, {de Simoni}, {Dean},
  {Dhital}, {Ealet}, {Ebelke}, {Edmondson}, {Eiting}, {Escoffier}, {Esposito},
  {Evans}, {Fan}, {Femen{\'\i}a Castell{\'a}}, {Dutra Ferreira}, {Fitzgerald},
  {Fleming}, {Font-Ribera}, {Ford}, {Frinchaboy}, {Garc{\'\i}a P{\'e}rez},
  {Gaudi}, {Ge}, {Ghezzi}, {Gillespie}, {Gilmore}, {Girardi}, {Gott}, {Gould},
  {Grebel}, {Gunn}, {Hamilton}, {Harding}, {Harris}, {Hawley}, {Hearty},
  {Hennawi}, {Gonz{\'a}lez Hern{\'a}ndez}, {Ho}, {Hogg}, {Holtzman},
  {Honscheid}, {Inada}, {Ivans}, {Jiang}, {Jiang}, {Johnson}, {Jordan},
  {Jordan}, {Kauffmann}, {Kazin}, {Kirkby}, {Klaene}, {Knapp}, {Kneib},
  {Kochanek}, {Koesterke}, {Kollmeier}, {Kron}, {Lampeitl}, {Lang}, {Lawler},
  {Le Goff}, {Lee}, {Lee}, {Leisenring}, {Lin}, {Liu}, {Long}, {Loomis},
  {Lucatello}, {Lundgren}, {Lupton}, {Ma}, {Ma}, {MacDonald}, {Mack},
  {Mahadevan}, {Maia}, {Majewski}, {Makler}, {Malanushenko}, {Malanushenko},
  {Mandelbaum}, {Maraston}, {Margala}, {Maseman}, {Masters}, {McBride},
  {McDonald}, {McGreer}, {McMahon}, {Mena Requejo}, {M{\'e}nard},
  {Miralda-Escud{\'e}}, {Morrison}, {Mullally}, {Muna}, {Murayama}, {Myers},
  {Naugle}, {Neto}, {Nguyen}, {Nichol}, {Nidever}, {O'Connell}, {Ogando},
  {Olmstead}, {Oravetz}, {Padmanabhan}, {Paegert}, {Palanque-Delabrouille},
  {Pan}, {Pandey}, {Parejko}, {P{\^a}ris}, {Pellegrini}, {Pepper}, {Percival},
  {Petitjean}, {Pfaffenberger}, {Pforr}, {Phleps}, {Pichon}, {Pieri}, {Prada},
  {Price-Whelan}, {Raddick}, {Ramos}, {Reid}, {Reyle}, {Rich}, {Richards},
  {Rieke}, {Rieke}, {Rix}, {Robin}, {Rocha-Pinto}, {Rockosi}, {Roe},
  {Rollinde}, {Ross}, {Ross}, {Rossetto}, {S{\'a}nchez}, {Santiago}, {Sayres},
  {Schiavon}, {Schlegel}, {Schlesinger}, {Schmidt}, {Schneider}, {Sellgren},
  {Shelden}, {Sheldon}, {Shetrone}, {Shu}, {Silverman}, {Simmerer}, {Simmons},
  {Sivarani}, {Skrutskie}, {Slosar}, {Smee}, {Smith}, {Snedden}, {Stassun},
  {Steele}, {Steinmetz}, {Stockett}, {Stollberg}, {Strauss}, {Szalay},
  {Tanaka}, {Thakar}, {Thomas}, {Tinker}, {Tofflemire}, {Tojeiro}, {Tremonti},
  {Vargas Maga{\~n}a}, {Verde}, {Vogt}, {Wake}, {Wan}, {Wang}, {Weaver},
  {White}, {White}, {Wilson}, {Wisniewski}, {Wood-Vasey}, {Yanny}, {Yasuda},
  {Y{\`e}che}, {York}, {Young}, {Zasowski}, {Zehavi}, \&
  {Zhao}}]{Eisenstein_2011}
{Eisenstein}, D.~J., {Weinberg}, D.~H., {Agol}, E., {et~al.} 2011,
  \href{http://dx.doi.org/10.1088/0004-6256/142/3/72}{\JournalTitle{\aj}, 142,
  72}

\bibitem[{{El{\'\i}asd{\'o}ttir} {et~al.}(2007){El{\'\i}asd{\'o}ttir},
  {Limousin}, {Richard}, {Hjorth}, {Kneib}, {Natarajan}, {Pedersen}, {Jullo},
  \& {Paraficz}}]{Eliasdottir_2007}
{El{\'\i}asd{\'o}ttir}, {\'A}., {Limousin}, M., {Richard}, J., {et~al.} 2007,
  \JournalTitle{arXiv e-prints}, arXiv:0710.5636

\bibitem[{{Fagioli} {et~al.}(2016){Fagioli}, {Carollo}, {Renzini}, {Lilly},
  {Onodera}, \& {Tacchella}}]{Fagioli_2016}
{Fagioli}, M., {Carollo}, C.~M., {Renzini}, A., {et~al.} 2016,
  \href{http://dx.doi.org/10.3847/0004-637X/831/2/173}{\JournalTitle{\apj},
  831, 173}

\bibitem[{{Fan} {et~al.}(2010){Fan}, {Lapi}, {Bressan}, {Bernardi}, {De Zotti},
  \& {Danese}}]{Fan_2010}
{Fan}, L., {Lapi}, A., {Bressan}, A., {et~al.} 2010,
  \href{http://dx.doi.org/10.1088/0004-637X/718/2/1460}{\JournalTitle{\apj},
  718, 1460}

\bibitem[{{Feldmann} {et~al.}(2016){Feldmann}, {Hopkins}, {Quataert},
  {Faucher-Gigu{\`e}re}, \& {Kere{\v{s}}}}]{Feldman_2016}
{Feldmann}, R., {Hopkins}, P.~F., {Quataert}, E., {Faucher-Gigu{\`e}re}, C.-A.,
  \& {Kere{\v{s}}}, D. 2016,
  \href{http://dx.doi.org/10.1093/mnrasl/slw014}{\JournalTitle{\mnras}, 458,
  L14}

\bibitem[{{Ferr{\'e}-Mateu} {et~al.}(2017){Ferr{\'e}-Mateu}, {Trujillo},
  {Mart{\'\i}n-Navarro}, {Vazdekis}, {Mezcua}, {Balcells}, \&
  {Dom{\'\i}nguez}}]{FerreMateu_2017}
{Ferr{\'e}-Mateu}, A., {Trujillo}, I., {Mart{\'\i}n-Navarro}, I., {et~al.}
  2017, \href{http://dx.doi.org/10.1093/mnras/stx171}{\JournalTitle{\mnras},
  467, 1929}

\bibitem[{{Foreman-Mackey} {et~al.}(2013){Foreman-Mackey}, {Hogg}, {Lang}, \&
  {Goodman}}]{emceehammer}
{Foreman-Mackey}, D., {Hogg}, D.~W., {Lang}, D., \& {Goodman}, J. 2013,
  \href{http://dx.doi.org/10.1086/670067}{\JournalTitle{\pasp}, 125, 306}

\bibitem[{{Garg} {et~al.}(2007){Garg}, {Stubbs}, {Challis}, {Wood-Vasey},
  {Blondin}, {Huber}, {Cook}, {Nikolaev}, {Rest}, {Smith}, {Olsen}, {Suntzeff},
  {Aguilera}, {Prieto}, {Becker}, {Miceli}, {Miknaitis}, {Clocchiatti},
  {Minniti}, {Morelli}, \& {Welch}}]{Garg_2007}
{Garg}, A., {Stubbs}, C.~W., {Challis}, P., {et~al.} 2007,
  \href{http://dx.doi.org/10.1086/510118}{\JournalTitle{\aj}, 133, 403}

\bibitem[{{Gladders} \& {Yee}(2000)}]{Gladders2000}
{Gladders}, M.~D., \& {Yee}, H.~K.~C. 2000,
  \href{http://dx.doi.org/10.1086/301557}{\JournalTitle{\aj}, 120, 2148}

\bibitem[{{Hao} {et~al.}(2010){Hao}, {McKay}, {Koester}, {Rykoff}, {Rozo},
  {Annis}, {Wechsler}, {Evrard}, {Siegel}, {Becker}, {Busha}, {Gerdes},
  {Johnston}, \& {Sheldon}}]{Hao_2010}
{Hao}, J., {McKay}, T.~A., {Koester}, B.~P., {et~al.} 2010,
  \href{http://dx.doi.org/10.1088/0067-0049/191/2/254}{\JournalTitle{\apjs},
  191, 254}

\bibitem[{Harris {et~al.}(2020)Harris, Millman, van~der Walt, Gommers,
  Virtanen, Cournapeau, Wieser, Taylor, Berg, Smith, Kern, Picus, Hoyer, van
  Kerkwijk, Brett, Haldane, del R{\'{i}}o, Wiebe, Peterson,
  G{\'{e}}rard-Marchant, Sheppard, Reddy, Weckesser, Abbasi, Gohlke, \&
  Oliphant}]{numpy}
Harris, C.~R., Millman, K.~J., van~der Walt, S.~J., {et~al.} 2020,
  \href{http://dx.doi.org/10.1038/s41586-020-2649-2}{\JournalTitle{Nature},
  585, 357}

\bibitem[{{Hilton} {et~al.}(2021){Hilton}, {Sif{\'o}n}, {Naess},
  {Madhavacheril}, {Oguri}, {Rozo}, {Rykoff}, {Abbott}, {Adhikari}, {Aguena},
  {Aiola}, {Allam}, {Amodeo}, {Amon}, {Annis}, {Ansarinejad}, {Aros-Bunster},
  {Austermann}, {Avila}, {Bacon}, {Battaglia}, {Beall}, {Becker}, {Bernstein},
  {Bertin}, {Bhandarkar}, {Bhargava}, {Bond}, {Brooks}, {Burke}, {Calabrese},
  {Carrasco Kind}, {Carretero}, {Choi}, {Choi}, {Conselice}, {da Costa},
  {Costanzi}, {Crichton}, {Crowley}, {D{\"u}nner}, {Denison}, {Devlin},
  {Dicker}, {Diehl}, {Dietrich}, {Doel}, {Duff}, {Duivenvoorden}, {Dunkley},
  {Everett}, {Ferraro}, {Ferrero}, {Fert{\'e}}, {Flaugher}, {Frieman},
  {Gallardo}, {Garc{\'\i}a-Bellido}, {Gaztanaga}, {Gerdes}, {Giles}, {Golec},
  {Gralla}, {Grandis}, {Gruen}, {Gruendl}, {Gschwend}, {Gutierrez}, {Han},
  {Hartley}, {Hasselfield}, {Hill}, {Hilton}, {Hincks}, {Hinton}, {Ho},
  {Honscheid}, {Hoyle}, {Hubmayr}, {Huffenberger}, {Hughes}, {Jaelani}, {Jain},
  {James}, {Jeltema}, {Kent}, {Knowles}, {Koopman}, {Kuehn}, {Lahav}, {Lima},
  {Lin}, {Lokken}, {Loubser}, {MacCrann}, {Maia}, {Marriage}, {Martin},
  {McMahon}, {Melchior}, {Menanteau}, {Miquel}, {Miyatake}, {Moodley},
  {Morgan}, {Mroczkowski}, {Nati}, {Newburgh}, {Niemack}, {Nishizawa},
  {Ogando}, {Orlowski-Scherer}, {Page}, {Palmese}, {Partridge},
  {Paz-Chinch{\'o}n}, {Phakathi}, {Plazas}, {Robertson}, {Romer}, {Carnero
  Rosell}, {Salatino}, {Sanchez}, {Schaan}, {Schillaci}, {Sehgal}, {Serrano},
  {Shin}, {Simon}, {Smith}, {Soares-Santos}, {Spergel}, {Staggs}, {Storer},
  {Suchyta}, {Swanson}, {Tarle}, {Thomas}, {To}, {Trac}, {Ullom}, {Vale}, {Van
  Lanen}, {Vavagiakis}, {De Vicente}, {Wilkinson}, {Wollack}, {Xu}, \&
  {Zhang}}]{Hilton_2021}
{Hilton}, M., {Sif{\'o}n}, C., {Naess}, S., {et~al.} 2021,
  \href{http://dx.doi.org/10.3847/1538-4365/abd023}{\JournalTitle{\apjs}, 253,
  3}

\bibitem[{Hilz {et~al.}(2013)Hilz, Naab, \& Ostriker}]{Hilz_2013}
Hilz, M., Naab, T., \& Ostriker, J.~P. 2013,
  \href{http://dx.doi.org/10.1093/mnras/sts501}{\JournalTitle{Monthly Notices
  of the Royal Astronomical Society}, 429, 2924}

\bibitem[{{Hinshaw} {et~al.}(2013){Hinshaw}, {Larson}, {Komatsu}, {Spergel},
  {Bennett}, {Dunkley}, {Nolta}, {Halpern}, {Hill}, {Odegard}, {Page}, {Smith},
  {Weiland}, {Gold}, {Jarosik}, {Kogut}, {Limon}, {Meyer}, {Tucker}, {Wollack},
  \& {Wright}}]{Hinshaw_2013}
{Hinshaw}, G., {Larson}, D., {Komatsu}, E., {et~al.} 2013,
  \href{http://dx.doi.org/10.1088/0067-0049/208/2/19}{\JournalTitle{\apjs},
  208, 19}

\bibitem[{{Hopkins} {et~al.}(2009){Hopkins}, {Bundy}, {Murray}, {Quataert},
  {Lauer}, \& {Ma}}]{Hopkins_2009}
{Hopkins}, P.~F., {Bundy}, K., {Murray}, N., {et~al.} 2009,
  \href{http://dx.doi.org/10.1111/j.1365-2966.2009.15062.x}{\JournalTitle{\mnras},
  398, 898}

\bibitem[{{Hsu} {et~al.}(2014){Hsu}, {Stockton}, \& {Shih}}]{Hsu_2014}
{Hsu}, L.-Y., {Stockton}, A., \& {Shih}, H.-Y. 2014,
  \href{http://dx.doi.org/10.1088/0004-637X/796/2/92}{\JournalTitle{\apj}, 796,
  92}

\bibitem[{{Huang} {et~al.}(2020){Huang}, {Storfer}, {Ravi}, {Pilon}, {Domingo},
  {Schlegel}, {Bailey}, {Dey}, {Gupta}, {Herrera}, {Juneau}, {Landriau},
  {Lang}, {Meisner}, {Moustakas}, {Myers}, {Schlafly}, {Valdes}, {Weaver},
  {Yang}, \& {Y{\`e}che}}]{Huang_2020}
{Huang}, X., {Storfer}, C., {Ravi}, V., {et~al.} 2020,
  \href{http://dx.doi.org/10.3847/1538-4357/ab7ffb}{\JournalTitle{\apj}, 894,
  78}

\bibitem[{{Huang} {et~al.}(2021){Huang}, {Storfer}, {Gu}, {Ravi}, {Pilon},
  {Sheu}, {Venguswamy}, {Banka}, {Dey}, {Landriau}, {Lang}, {Meisner},
  {Moustakas}, {Myers}, {Sajith}, {Schlafly}, \& {Schlegel}}]{Huang_2021}
{Huang}, X., {Storfer}, C., {Gu}, A., {et~al.} 2021,
  \href{http://dx.doi.org/10.3847/1538-4357/abd62b}{\JournalTitle{\apj}, 909,
  27}

\bibitem[{Hunter(2007)}]{matplotlib}
Hunter, J.~D. 2007,
  \href{http://dx.doi.org/10.1109/MCSE.2007.55}{\JournalTitle{Computing in
  Science Engineering}, 9, 90}

\bibitem[{{Jafariyazani} {et~al.}(2020){Jafariyazani}, {Newman}, {Mobasher},
  {Belli}, {Ellis}, \& {Patel}}]{Jafariyazani_2020}
{Jafariyazani}, M., {Newman}, A.~B., {Mobasher}, B., {et~al.} 2020,
  \href{http://dx.doi.org/10.3847/2041-8213/aba11c}{\JournalTitle{\apjl}, 897,
  L42}

\bibitem[{Johnson \& Leja(2017)}]{prospector}
Johnson, B., \& Leja, J. 2017, bd-j/prospector: Initial release

\bibitem[{{Joye} \& {Mandel}(2003)}]{Joye_2003}
{Joye}, W.~A., \& {Mandel}, E. 2003, in Astronomical Society of the Pacific
  Conference Series, Vol. 295, Astronomical Data Analysis Software and Systems
  XII, ed. H.~E. {Payne}, R.~I. {Jedrzejewski}, \& R.~N. {Hook}, 489

\bibitem[{{Jullo} {et~al.}(2007){Jullo}, {Kneib}, {Limousin},
  {El{\'\i}asd{\'o}ttir}, {Marshall}, \& {Verdugo}}]{Jullo_2007}
{Jullo}, E., {Kneib}, J.~P., {Limousin}, M., {et~al.} 2007,
  \href{http://dx.doi.org/10.1088/1367-2630/9/12/447}{\JournalTitle{New Journal
  of Physics}, 9, 447}

\bibitem[{{Khullar} {et~al.}(2021){Khullar}, {Gozman}, {Lin}, {Martinez},
  {Matthews Acu{\~n}a}, {Medina}, {Merz}, {Sanchez}, {Sisco}, {Kavin Stein},
  {Sukay}, {Tavangar}, {Bayliss}, {Bleem}, {Brownsberger}, {Dahle}, {Florian},
  {Gladders}, {Mahler}, {Rigby}, {Sharon}, \& {Stark}}]{Khullar_2021}
{Khullar}, G., {Gozman}, K., {Lin}, J.~J., {et~al.} 2021,
  \href{http://dx.doi.org/10.3847/1538-4357/abcb86}{\JournalTitle{\apj}, 906,
  107}

\bibitem[{{Kriek} \& {Conroy}(2013)}]{Kriek_2013}
{Kriek}, M., \& {Conroy}, C. 2013,
  \href{http://dx.doi.org/10.1088/2041-8205/775/1/L16}{\JournalTitle{\apjl},
  775, L16}

\bibitem[{{Kriek} {et~al.}(2008){Kriek}, {van Dokkum}, {Franx}, {Illingworth},
  {Marchesini}, {Quadri}, {Rudnick}, {Taylor}, {F{\"o}rster Schreiber},
  {Gawiser}, {Labb{\'e}}, {Lira}, \& {Wuyts}}]{Kriek_2008}
{Kriek}, M., {van Dokkum}, P.~G., {Franx}, M., {et~al.} 2008,
  \href{http://dx.doi.org/10.1086/528945}{\JournalTitle{\apj}, 677, 219}

\bibitem[{{Lanusse} {et~al.}(2018){Lanusse}, {Ma}, {Li}, {Collett}, {Li},
  {Ravanbakhsh}, {Mandelbaum}, \& {P{\'o}czos}}]{Lanusse_2018}
{Lanusse}, F., {Ma}, Q., {Li}, N., {et~al.} 2018,
  \href{http://dx.doi.org/10.1093/mnras/stx1665}{\JournalTitle{\mnras}, 473,
  3895}

\bibitem[{{Leja} {et~al.}(2019){Leja}, {Carnall}, {Johnson}, {Conroy}, \&
  {Speagle}}]{Leja_2019}
{Leja}, J., {Carnall}, A.~C., {Johnson}, B.~D., {Conroy}, C., \& {Speagle},
  J.~S. 2019,
  \href{http://dx.doi.org/10.3847/1538-4357/ab133c}{\JournalTitle{\apj}, 876,
  3}

\bibitem[{{Leja} {et~al.}(2017){Leja}, {Johnson}, {Conroy}, {van Dokkum}, \&
  {Byler}}]{Leja_2017}
{Leja}, J., {Johnson}, B.~D., {Conroy}, C., {van Dokkum}, P.~G., \& {Byler}, N.
  2017, \href{http://dx.doi.org/10.3847/1538-4357/aa5ffe}{\JournalTitle{\apj},
  837, 170}

\bibitem[{{Man} {et~al.}(2021){Man}, {Zabl}, {Brammer}, {Richard}, {Toft},
  {Stockmann}, {Gallazzi}, {Zibetti}, \& {Ebeling}}]{Man_2021}
{Man}, A. W.~S., {Zabl}, J., {Brammer}, G.~B., {et~al.} 2021,
  \href{http://dx.doi.org/10.3847/1538-4357/ac0ae3}{\JournalTitle{\apj}, 919,
  20}

\bibitem[{{Miknaitis} {et~al.}(2007){Miknaitis}, {Pignata}, {Rest},
  {Wood-Vasey}, {Blondin}, {Challis}, {Smith}, {Stubbs}, {Suntzeff}, {Foley},
  {Matheson}, {Tonry}, {Aguilera}, {Blackman}, {Becker}, {Clocchiatti},
  {Covarrubias}, {Davis}, {Filippenko}, {Garg}, {Garnavich}, {Hicken}, {Jha},
  {Krisciunas}, {Kirshner}, {Leibundgut}, {Li}, {Miceli}, {Narayan}, {Prieto},
  {Riess}, {Salvo}, {Schmidt}, {Sollerman}, {Spyromilio}, \&
  {Zenteno}}]{Miknaitis_2007}
{Miknaitis}, G., {Pignata}, G., {Rest}, A., {et~al.} 2007,
  \href{http://dx.doi.org/10.1086/519986}{\JournalTitle{\apj}, 666, 674}

\bibitem[{{Muzzin} {et~al.}(2013){Muzzin}, {Marchesini}, {Stefanon}, {Franx},
  {McCracken}, {Milvang-Jensen}, {Dunlop}, {Fynbo}, {Brammer}, {Labb{\'e}}, \&
  {van Dokkum}}]{Muzzin_2013}
{Muzzin}, A., {Marchesini}, D., {Stefanon}, M., {et~al.} 2013,
  \href{http://dx.doi.org/10.1088/0004-637X/777/1/18}{\JournalTitle{\apj}, 777,
  18}

\bibitem[{{Naab} {et~al.}(2009){Naab}, {Johansson}, \& {Ostriker}}]{Naab_2009}
{Naab}, T., {Johansson}, P.~H., \& {Ostriker}, J.~P. 2009,
  \href{http://dx.doi.org/10.1088/0004-637X/699/2/L178}{\JournalTitle{\apjl},
  699, L178}

\bibitem[{{Newman} {et~al.}(2018){Newman}, {Belli}, {Ellis}, \&
  {Patel}}]{Newman_2018}
{Newman}, A.~B., {Belli}, S., {Ellis}, R.~S., \& {Patel}, S.~G. 2018,
  \href{http://dx.doi.org/10.3847/1538-4357/aacd4d}{\JournalTitle{\apj}, 862,
  125}

\bibitem[{{Newman} {et~al.}(2012){Newman}, {Ellis}, {Bundy}, \&
  {Treu}}]{Newman_2012}
{Newman}, A.~B., {Ellis}, R.~S., {Bundy}, K., \& {Treu}, T. 2012,
  \href{http://dx.doi.org/10.1088/0004-637X/746/2/162}{\JournalTitle{\apj},
  746, 162}

\bibitem[{{Newman} {et~al.}(2010){Newman}, {Ellis}, {Treu}, \&
  {Bundy}}]{Newman_2010}
{Newman}, A.~B., {Ellis}, R.~S., {Treu}, T., \& {Bundy}, K. 2010,
  \href{http://dx.doi.org/10.1088/2041-8205/717/2/L103}{\JournalTitle{\apjl},
  717, L103}

\bibitem[{{Nipoti} {et~al.}(2012){Nipoti}, {Treu}, {Leauthaud}, {Bundy},
  {Newman}, \& {Auger}}]{Nipoti_2012}
{Nipoti}, C., {Treu}, T., {Leauthaud}, A., {et~al.} 2012,
  \href{http://dx.doi.org/10.1111/j.1365-2966.2012.20749.x}{\JournalTitle{\mnras},
  422, 1714}

\bibitem[{{Oldham} {et~al.}(2017){Oldham}, {Auger}, {Fassnacht}, {Treu},
  {Brewer}, {Koopmans}, {Lagattuta}, {Marshall}, {McKean}, \&
  {Vegetti}}]{Oldham_2017}
{Oldham}, L., {Auger}, M.~W., {Fassnacht}, C.~D., {et~al.} 2017,
  \href{http://dx.doi.org/10.1093/mnras/stw2832}{\JournalTitle{\mnras}, 465,
  3185}

\bibitem[{{Peng} {et~al.}(2002){Peng}, {Ho}, {Impey}, \& {Rix}}]{Peng_2002}
{Peng}, C.~Y., {Ho}, L.~C., {Impey}, C.~D., \& {Rix}, H.-W. 2002,
  \href{http://dx.doi.org/10.1086/340952}{\JournalTitle{\aj}, 124, 266}

\bibitem[{{Peng} {et~al.}(2010){Peng}, {Ho}, {Impey}, \& {Rix}}]{Peng_2010}
---. 2010,
  \href{http://dx.doi.org/10.1088/0004-6256/139/6/2097}{\JournalTitle{\aj},
  139, 2097}

\bibitem[{{Persson} {et~al.}(2008){Persson}, {Barkhouser}, {Birk}, {Hammond},
  {Harding}, {Koch}, {Marshall}, {McCarthy}, {Murphy}, {Orndorff},
  {Scharfstein}, {Shectman}, {Smee}, \& {Uomoto}}]{Persson_2008}
{Persson}, S.~E., {Barkhouser}, R., {Birk}, C., {et~al.} 2008,
  \href{http://dx.doi.org/10.1117/12.790015}{in Society of Photo-Optical
  Instrumentation Engineers (SPIE) Conference Series, Vol. 7014, Ground-based
  and Airborne Instrumentation for Astronomy II}, 70142V

\bibitem[{{Rest} {et~al.}(2005){Rest}, {Stubbs}, {Becker}, {Miknaitis},
  {Miceli}, {Covarrubias}, {Hawley}, {Smith}, {Suntzeff}, {Olsen}, {Prieto},
  {Hiriart}, {Welch}, {Cook}, {Nikolaev}, {Huber}, {Prochtor}, {Clocchiatti},
  {Minniti}, {Garg}, {Challis}, {Keller}, \& {Schmidt}}]{Rest_2005}
{Rest}, A., {Stubbs}, C., {Becker}, A.~C., {et~al.} 2005,
  \href{http://dx.doi.org/10.1086/497060}{\JournalTitle{\apj}, 634, 1103}

\bibitem[{{Rykoff} {et~al.}(2014){Rykoff}, {Rozo}, {Busha}, {Cunha},
  {Finoguenov}, {Evrard}, {Hao}, {Koester}, {Leauthaud}, {Nord}, {Pierre},
  {Reddick}, {Sadibekova}, {Sheldon}, \& {Wechsler}}]{Rykoff_2014}
{Rykoff}, E.~S., {Rozo}, E., {Busha}, M.~T., {et~al.} 2014,
  \href{http://dx.doi.org/10.1088/0004-637X/785/2/104}{\JournalTitle{\apj},
  785, 104}

\bibitem[{{Salim} {et~al.}(2018){Salim}, {Boquien}, \& {Lee}}]{Salim_2018}
{Salim}, S., {Boquien}, M., \& {Lee}, J.~C. 2018,
  \href{http://dx.doi.org/10.3847/1538-4357/aabf3c}{\JournalTitle{\apj}, 859,
  11}

\bibitem[{{Schlafly} \& {Finkbeiner}(2011)}]{Schlafly_2011}
{Schlafly}, E.~F., \& {Finkbeiner}, D.~P. 2011,
  \href{http://dx.doi.org/10.1088/0004-637X/737/2/103}{\JournalTitle{\apj},
  737, 103}

\bibitem[{Sharon {et~al.}(2020)Sharon, Bayliss, Dahle, Dunham, Florian,
  Gladders, Johnson, Mahler, Paterno-Mahler, Rigby, Whitaker, Akhshik, Koester,
  Murray, Gonz{\'{a}}lez, \& Wuyts}]{Sharon_2020}
Sharon, K., Bayliss, M.~B., Dahle, H., {et~al.} 2020,
  \href{http://dx.doi.org/10.3847/1538-4365/ab5f13}{\JournalTitle{The
  Astrophysical Journal Supplement Series}, 247, 12}

\bibitem[{{Skrutskie} {et~al.}(2006){Skrutskie}, {Cutri}, {Stiening},
  {Weinberg}, {Schneider}, {Carpenter}, {Beichman}, {Capps}, {Chester},
  {Elias}, {Huchra}, {Liebert}, {Lonsdale}, {Monet}, {Price}, {Seitzer},
  {Jarrett}, {Kirkpatrick}, {Gizis}, {Howard}, {Evans}, {Fowler}, {Fullmer},
  {Hurt}, {Light}, {Kopan}, {Marsh}, {McCallon}, {Tam}, {Van Dyk}, \&
  {Wheelock}}]{Skrutskie_2006}
{Skrutskie}, M.~F., {Cutri}, R.~M., {Stiening}, R., {et~al.} 2006,
  \href{http://dx.doi.org/10.1086/498708}{\JournalTitle{\aj}, 131, 1163}

\bibitem[{{Stockton} {et~al.}(2010){Stockton}, {Shih}, \&
  {Larson}}]{Stockton_2010}
{Stockton}, A., {Shih}, H.-Y., \& {Larson}, K. 2010,
  \href{http://dx.doi.org/10.1088/2041-8205/709/1/L58}{\JournalTitle{\apjl},
  709, L58}

\bibitem[{{Stockton} {et~al.}(2014){Stockton}, {Shih}, {Larson}, \&
  {Mann}}]{Stockton_2014}
{Stockton}, A., {Shih}, H.-Y., {Larson}, K., \& {Mann}, A.~W. 2014,
  \href{http://dx.doi.org/10.1088/0004-637X/780/2/134}{\JournalTitle{\apj},
  780, 134}

\bibitem[{{Strauss} {et~al.}(2002){Strauss}, {Weinberg}, {Lupton}, {Narayanan},
  {Annis}, {Bernardi}, {Blanton}, {Burles}, {Connolly}, {Dalcanton}, {Doi},
  {Eisenstein}, {Frieman}, {Fukugita}, {Gunn}, {Ivezi{\'c}}, {Kent}, {Kim},
  {Knapp}, {Kron}, {Munn}, {Newberg}, {Nichol}, {Okamura}, {Quinn}, {Richmond},
  {Schlegel}, {Shimasaku}, {SubbaRao}, {Szalay}, {Vanden Berk}, {Vogeley},
  {Yanny}, {Yasuda}, {York}, \& {Zehavi}}]{Strauss_2002}
{Strauss}, M.~A., {Weinberg}, D.~H., {Lupton}, R.~H., {et~al.} 2002,
  \href{http://dx.doi.org/10.1086/342343}{\JournalTitle{\aj}, 124, 1810}

\bibitem[{{Tody}(1986)}]{Tody1986}
{Tody}, D. 1986, \href{http://dx.doi.org/10.1117/12.968154}{in Society of
  Photo-Optical Instrumentation Engineers (SPIE) Conference Series, Vol. 627,
  Instrumentation in astronomy VI, ed. D.~L. {Crawford}}, 733

\bibitem[{{Tody}(1993)}]{Tody1993}
{Tody}, D. 1993, in Astronomical Society of the Pacific Conference Series,
  Vol.~52, Astronomical Data Analysis Software and Systems II, ed. R.~J.
  {Hanisch}, R.~J.~V. {Brissenden}, \& J.~{Barnes}, 173

\bibitem[{{Trujillo} {et~al.}(2007){Trujillo}, {Conselice}, {Bundy}, {Cooper},
  {Eisenhardt}, \& {Ellis}}]{Trujillo_2007}
{Trujillo}, I., {Conselice}, C.~J., {Bundy}, K., {et~al.} 2007,
  \href{http://dx.doi.org/10.1111/j.1365-2966.2007.12388.x}{\JournalTitle{\mnras},
  382, 109}

\bibitem[{{Trujillo} {et~al.}(2014){Trujillo}, {Ferr{\'e}-Mateu}, {Balcells},
  {Vazdekis}, \& {S{\'a}nchez-Bl{\'a}zquez}}]{Trujillo_2014}
{Trujillo}, I., {Ferr{\'e}-Mateu}, A., {Balcells}, M., {Vazdekis}, A., \&
  {S{\'a}nchez-Bl{\'a}zquez}, P. 2014,
  \href{http://dx.doi.org/10.1088/2041-8205/780/2/L20}{\JournalTitle{\apjl},
  780, L20}

\bibitem[{{Trujillo} {et~al.}(2006){Trujillo}, {Feulner}, {Goranova}, {Hopp},
  {Longhetti}, {Saracco}, {Bender}, {Braito}, {Della Ceca}, {Drory},
  {Mannucci}, \& {Severgnini}}]{Trujillo_2006}
{Trujillo}, I., {Feulner}, G., {Goranova}, Y., {et~al.} 2006,
  \href{http://dx.doi.org/10.1111/j.1745-3933.2006.00238.x}{\JournalTitle{\mnras},
  373, L36}

\bibitem[{{van der Wel} {et~al.}(2014){van der Wel}, {Franx}, {van Dokkum},
  {Skelton}, {Momcheva}, {Whitaker}, {Brammer}, {Bell}, {Rix}, {Wuyts},
  {Ferguson}, {Holden}, {Barro}, {Koekemoer}, {Chang}, {McGrath},
  {H{\"a}ussler}, {Dekel}, {Behroozi}, {Fumagalli}, {Leja}, {Lundgren},
  {Maseda}, {Nelson}, {Wake}, {Patel}, {Labb{\'e}}, {Faber}, {Grogin}, \&
  {Kocevski}}]{vanderWel_2014}
{van der Wel}, A., {Franx}, M., {van Dokkum}, P.~G., {et~al.} 2014,
  \href{http://dx.doi.org/10.1088/0004-637X/788/1/28}{\JournalTitle{\apj}, 788,
  28}

\bibitem[{{van der Wel} {et~al.}(2016){van der Wel}, {Noeske}, {Bezanson},
  {Pacifici}, {Gallazzi}, {Franx}, {Mu{\~n}oz-Mateos}, {Bell}, {Brammer},
  {Charlot}, {Chauk{\'e}}, {Labb{\'e}}, {Maseda}, {Muzzin}, {Rix}, {Sobral},
  {van de Sande}, {van Dokkum}, {Wild}, \& {Wolf}}]{vanderWel_2016}
{van der Wel}, A., {Noeske}, K., {Bezanson}, R., {et~al.} 2016,
  \href{http://dx.doi.org/10.3847/0067-0049/223/2/29}{\JournalTitle{\apjs},
  223, 29}

\bibitem[{{van Dokkum} {et~al.}(2008){van Dokkum}, {Franx}, {Kriek}, {Holden},
  {Illingworth}, {Magee}, {Bouwens}, {Marchesini}, {Quadri}, {Rudnick},
  {Taylor}, \& {Toft}}]{VanDokkum_2008}
{van Dokkum}, P.~G., {Franx}, M., {Kriek}, M., {et~al.} 2008,
  \href{http://dx.doi.org/10.1086/587874}{\JournalTitle{\apjl}, 677, L5}

\bibitem[{{van Dokkum} {et~al.}(2010){van Dokkum}, {Whitaker}, {Brammer},
  {Franx}, {Kriek}, {Labb{\'e}}, {Marchesini}, {Quadri}, {Bezanson},
  {Illingworth}, {Muzzin}, {Rudnick}, {Tal}, \& {Wake}}]{vanDokkum_2010}
{van Dokkum}, P.~G., {Whitaker}, K.~E., {Brammer}, G., {et~al.} 2010,
  \href{http://dx.doi.org/10.1088/0004-637X/709/2/1018}{\JournalTitle{\apj},
  709, 1018}

\bibitem[{Virtanen {et~al.}(2020)Virtanen, Gommers, Oliphant, Haberland, Reddy,
  Cournapeau, Burovski, Peterson, Weckesser, Bright, {van der Walt}, Brett,
  Wilson, Millman, Mayorov, Nelson, Jones, Kern, Larson, Carey, Polat, Feng,
  Moore, {VanderPlas}, Laxalde, Perktold, Cimrman, Henriksen, Quintero, Harris,
  Archibald, Ribeiro, Pedregosa, {van Mulbregt}, \& {SciPy 1.0
  Contributors}}]{scipy}
Virtanen, P., Gommers, R., Oliphant, T.~E., {et~al.} 2020,
  \href{http://dx.doi.org/10.1038/s41592-019-0686-2}{\JournalTitle{Nature
  Methods}, 17, 261}

\bibitem[{{Wellons} {et~al.}(2015){Wellons}, {Torrey}, {Ma}, {Rodriguez-Gomez},
  {Vogelsberger}, {Kriek}, {van Dokkum}, {Nelson}, {Genel}, {Pillepich},
  {Springel}, {Sijacki}, {Snyder}, {Nelson}, {Sales}, \&
  {Hernquist}}]{Wellons_2015}
{Wellons}, S., {Torrey}, P., {Ma}, C.-P., {et~al.} 2015,
  \href{http://dx.doi.org/10.1093/mnras/stv303}{\JournalTitle{\mnras}, 449,
  361}

\bibitem[{{Whitaker} {et~al.}(2012){Whitaker}, {Kriek}, {van Dokkum},
  {Bezanson}, {Brammer}, {Franx}, \& {Labb{\'e}}}]{Whitaker_2012}
{Whitaker}, K.~E., {Kriek}, M., {van Dokkum}, P.~G., {et~al.} 2012,
  \href{http://dx.doi.org/10.1088/0004-637X/745/2/179}{\JournalTitle{\apj},
  745, 179}

\bibitem[{{Whitaker} {et~al.}(2021){Whitaker}, {Williams}, {Mowla}, {Spilker},
  {Toft}, {Narayanan}, {Pope}, {Magdis}, {van Dokkum}, {Akhshik}, {Bezanson},
  {Brammer}, {Leja}, {Man}, {Nelson}, {Richard}, {Pacifici}, {Sharon}, \&
  {Valentino}}]{Whitaker_2021}
{Whitaker}, K.~E., {Williams}, C.~C., {Mowla}, L., {et~al.} 2021,
  \href{http://dx.doi.org/10.1038/s41586-021-03806-7}{\JournalTitle{\nat}, 597,
  485}

\bibitem[{{Wuyts} {et~al.}(2010){Wuyts}, {Barrientos}, {Gladders}, {Sharon},
  {Bayliss}, {Carrasco}, {Gilbank}, {Yee}, {Koester}, \&
  {Mu{\~n}oz}}]{Wuyts2010}
{Wuyts}, E., {Barrientos}, L.~F., {Gladders}, M.~D., {et~al.} 2010,
  \href{http://dx.doi.org/10.1088/0004-637X/724/2/1182}{\JournalTitle{\apj},
  724, 1182}

\bibitem[{{York} {et~al.}(2000){York}, {Adelman}, {Anderson}, {Anderson},
  {Annis}, {Bahcall}, {Bakken}, {Barkhouser}, {Bastian}, {Berman}, {Boroski},
  {Bracker}, {Briegel}, {Briggs}, {Brinkmann}, {Brunner}, {Burles}, {Carey},
  {Carr}, {Castander}, {Chen}, {Colestock}, {Connolly}, {Crocker}, {Csabai},
  {Czarapata}, {Davis}, {Doi}, {Dombeck}, {Eisenstein}, {Ellman}, {Elms},
  {Evans}, {Fan}, {Federwitz}, {Fiscelli}, {Friedman}, {Frieman}, {Fukugita},
  {Gillespie}, {Gunn}, {Gurbani}, {de Haas}, {Haldeman}, {Harris}, {Hayes},
  {Heckman}, {Hennessy}, {Hindsley}, {Holm}, {Holmgren}, {Huang}, {Hull},
  {Husby}, {Ichikawa}, {Ichikawa}, {Ivezi{\'c}}, {Kent}, {Kim}, {Kinney},
  {Klaene}, {Kleinman}, {Kleinman}, {Knapp}, {Korienek}, {Kron}, {Kunszt},
  {Lamb}, {Lee}, {Leger}, {Limmongkol}, {Lindenmeyer}, {Long}, {Loomis},
  {Loveday}, {Lucinio}, {Lupton}, {MacKinnon}, {Mannery}, {Mantsch}, {Margon},
  {McGehee}, {McKay}, {Meiksin}, {Merelli}, {Monet}, {Munn}, {Narayanan},
  {Nash}, {Neilsen}, {Neswold}, {Newberg}, {Nichol}, {Nicinski}, {Nonino},
  {Okada}, {Okamura}, {Ostriker}, {Owen}, {Pauls}, {Peoples}, {Peterson},
  {Petravick}, {Pier}, {Pope}, {Pordes}, {Prosapio}, {Rechenmacher}, {Quinn},
  {Richards}, {Richmond}, {Rivetta}, {Rockosi}, {Ruthmansdorfer}, {Sandford},
  {Schlegel}, {Schneider}, {Sekiguchi}, {Sergey}, {Shimasaku}, {Siegmund},
  {Smee}, {Smith}, {Snedden}, {Stone}, {Stoughton}, {Strauss}, {Stubbs},
  {SubbaRao}, {Szalay}, {Szapudi}, {Szokoly}, {Thakar}, {Tremonti}, {Tucker},
  {Uomoto}, {Vanden Berk}, {Vogeley}, {Waddell}, {Wang}, {Watanabe},
  {Weinberg}, {Yanny}, {Yasuda}, \& {SDSS Collaboration}}]{York_2000}
{York}, D.~G., {Adelman}, J., {Anderson}, John~E., J., {et~al.} 2000,
  \href{http://dx.doi.org/10.1086/301513}{\JournalTitle{\aj}, 120, 1579}

\end{thebibliography}

\pagebreak
\pagebreak

\end{document}